\documentclass[superscriptaddress,onecolumn]{revtex4}
\usepackage{graphicx,epsfig}
\usepackage{amsmath}
\usepackage {amssymb}
\usepackage{multirow}
\usepackage{float}
\usepackage{xcolor}
\providecommand{\abs}[1]{\lvert#1\rvert}

\makeindex

\begin{document}



\begin{center}
\large \bf {Nonlinear 
Scalarization of Schwarzschild Black Hole
 in Scalar-Torsion Teleparallel Gravity}
\end{center}


\author{P. A. Gonz\'{a}lez}
\email{pablo.gonzalez@udp.cl} \affiliation{Facultad de
Ingenier\'{i}a y Ciencias, Universidad Diego Portales, Avenida Ej\'{e}rcito
Libertador 441, Casilla 298-V, Santiago, Chile.}

\author{Eleftherios Papantonopoulos}
\email{lpapa@central.ntua.gr}
\affiliation{Physics Division, School of Applied Mathematical and Physical Sciences, National Technical University of Athens, 15780 Zografou Campus,
    Athens, Greece.}

\author{Joaqu\'in Robledo}
\email{joaquin.montenegro@userena.cl}

\author{Yerko V\'asquez}
\email{yvasquez@userena.cl}
\affiliation{Departamento de F\'isica, Facultad de Ciencias, Universidad de La Serena,\\
Avenida Cisternas 1200, La Serena, Chile.}


\begin{abstract}

We consider a scalar field coupled to the torsion in Teleparallel Gravity with a coupling function that does not allow for a tachyonic instability to occur, and it leads to the formation of new black holes with scalar hair.  
The scalarized black hole solutions are asymptotically flat, and they are studied under their thermodynamics, and we show that they are meta-stable solutions without phase transitions of first and second order.  Also, we demonstrate the existence of distinct branches of solutions classified by the number of nodes of the scalar field. For the branches analysed, the scalarized solutions have lower free energy than the Schwarzschild black hole; therefore, they are always thermodinamically preferred. Also, we show that the scalarized black holes are entropically favoured compared to the Schwarzschild solution.

\end{abstract}

\maketitle

\tableofcontents


\clearpage


\section{Introduction}

General Relativity (GR) is a  classical theory
of gravity which is constructed out of a unique spacetime metric along with its Levi-Civita connection, and it gives second order derivatives in the metric tensor. In this way it is a healthy physically acceptable field theory avoiding unphysical propagating modes, like ghosts \cite{chaostro} and moreover successfully  passes all local observational tests with high accuracy for both weak and strong gravity \cite{Will:2005va}. An important tensor besides curvature is the torsion which can be used for the construction of gravitational theories. In these theories the field equations arise by variation of two dynamical variables the vielbein and the connection. It was realized that imposing the teleparallelism condition, we can express the standard Einstein gravity \cite{Unzicker:2005in,TEGR,TEGR22,Hayashi:1979qx,JGPereira,Arcos:2004tzt,Maluf:2013gaa,Pereira:2013qza, Capozziello:2022zzh} by a different geometry described in terms of torsion. In this gravity theory instead of the Levi-Civita connection the Weitzenb$\ddot{o}$ck connection is used, which has vanishing curvature but non-vanishing antisymmetric piece which means non-vanishing torsion, and therefore the only dynamical variable that remains is the vielbein.

A generalization of the  teleparallelism theory is to assume an arbitrary connection of vanishing curvature and then everything is again expressed in terms of the torsion, but now both the vielbein and the connection are dynamical. In this  formulation, the theory is both Lorentz and diffeomorphism invariant, while the standard  formulation of the teleparallel theory lacks local Lorentz invariance. In the teleparallel formulation of GR, the Einstein gravity action consists of the torsion scalar $\mathcal{T}$, which is a specific combination of quadratic torsion scalars and contains up to first-order vielbein derivatives. Then, the Einstein gravitational field equations are arising by the variation of this action, and for this reason the theory is termed as Teleparallel Equivalent of General Relativity (TEGR) \cite{Unzicker:2005in,TEGR,TEGR22,Hayashi:1979qx,JGPereira,Arcos:2004tzt,Maluf:2013gaa,
Pereira:2013qza, Capozziello:2022zzh}.

An interesting question is if we wand to produce theories that can be differentiated  from GR, what kind of theories can be produced in which the torsion can give sizeable physical effects. One such approach was inspired from the $f(R)$ modifications of GR, in which starting from TEGR instead of GR, $f(\mathcal{T})$ extensions can be constructed. These gravitational modification do not coincide with $f(R)$ and can have local black hole solutions \cite{Wang:2011xf,Miao:2011ki,Gonzalez:2011dr, Capozziello:2012zj,Atazadeh:2012am,Calza:2023hhi}, and produce variable cosmological models \cite{Ferraro:2006jd,Linder:2010py, Li:2013xea,Kofinas:2014aka,Kofinas:2014daa}.

Another simple and viable modification of GR are the scalar-tensor theories \cite{Fujii}. If a scalar field coupled to gravity backreacts to the background metric, new hairy black hole solutions would be generated. One of the first hairy black hole was found in \cite{BBMB} in an asymptotically flat spacetime but it was shown in \cite{bronnikov} that it was unstable because the scalar field was divergent on the event horizon. However, introducing a cosmological constant making the space-time asymptotically AdS/dS, such an irregular behaviour of the scalar field  on the horizon was avoided. Then, the scalar field is regular on the horizon and hairy black hole solutions were found with all the possible divergence were hidden behind the horizon \cite{Martinez:1996gn,Banados:1992wn,Martinez:2002ru, Winstanley:2002jt,Kolyvaris:2010yyf,Charmousis:2014zaa}.

Recently there is a lot of activity studying gravitational theories in which matter \newline
parametrized by a scalar field is directly coupled to second order algebraic curvature invariants. In this case hairy black holes are obtained and scalar hair is maintained by the interaction with the spacetime curvature. One  curvature correction is the Gauss-Bonnet (GB) term  which is ghost-free but it becomes a topological term in four-dimensional spacetime, it is a topological term and therefore a scalar field in four dimensions has to couple to the GB term in order to have non-trivial dynamics. These theories have been studied  extensively in an attempt to evade  the no-hair theorems  and obtained hairy black hole solutions. In particular, for certain classes of the coupling function it was shown that we have spontaneous scalarization of black holes \cite{Doneva_2018a,Silva_2018,Chen:2006ge,Antoniou_2018,Antoniou_2018a}. It was found  that below a certain critical mass the Schwarzschild black hole becomes unstable in regions of strong curvature, and then when the scalar field backreacts to the metric, new branches of scalarized black holes develop. The scalarization is triggered by the tachyonic mass due to the coupling between a scalar field and the GB term. The spontaneous scalarization of Kerr black holes was also observed in the scalar-Gauss-Bonnet theory \cite{Cunha:2019dwb}. Also, the phenomena was studied considering the leading-order scalar self-interactions \cite{Macedo:2019sem}, and massive scalar fields \cite{Peng:2020znl}. 

Bekenstein in \cite{Bekenstein:1992pj} introduced  disformal transformations which  consist of the most general coupling constructed from the metric and the scalar field that respects causality and the weak equivalence principle. In \cite{Bettoni:2013diz} it was showed that the mathematical structure of Horndeski theory is preserved under the disformal transformations. Thus disformal transformations provide a natural generalization of conformal transformations.  In this context, a minimal model of a massless scalar-tensor theory was proposed and investigated on how the disformal coupling affects the spontaneous scalarization of slowly rotating neutron stars, given that,  for negative values of the disformal coupling parameter between the scalar field and matter, scalarization can be suppressed, while for large positive values of the disformal coupling parameter stellar models cannot be obtained. Also, it was shown that in a certain range of the theory's parameter space the universal relation largely deviates from that of GR, allowing, in principle, to probe the existence of spontaneous scalarization with future observations \cite{Minamitsuji:2016hkk}.

T. Damour and G. Esposito-Farese in \cite{Damour:1993hw} studied a canonical scalar-tensor theory with coupling to the matter Lagrangian via a conformally transformed metric $A^2(\phi)g_{\mu\nu}$. The scalar-tensor theory (with one massless scalar field) contains one arbitrary ``coupling function" $A(\phi)$ and it is given by the action
\begin{equation}
\label{actionD}
    S=\frac{1}{16\pi G_*}\int d^4xg_*^{1/2}[R_*-2g_*^{\mu\nu}\partial_\mu\phi\partial_\nu \phi]+S_m[\Psi_m, A^2(\phi)g^*_{\mu\nu}]\,,
\end{equation}
where $G_*$ denotes a bare gravitational coupling constant, and $R_*=g_*^{\mu\nu}R^*_{\mu\nu}$ the curvature scalar of the ``Einstein-metric" $g^*_{\mu\nu}$. The last term denotes the action of the matter, which is a functional of some matter variables, collectively denoted by $\Psi_m$, and of the (``Jordan-Fierz") metric $\bar{g}_{\mu\nu}=A^2(\phi)g^*_{\mu\nu}$. Here, the gravity is mediated by one or several long-range scalar fields in addition to the usual Einstein's theory, and it can be seen as a mechanism for local modification of gravity. It is worth mentioning that Kaluza-Klein, supergravity, and superstring theories naturally give rise to massless scalar fields coupled to matter.

Bergmann in \cite{Dcited} proposed scalar-tensor theories, by  considering that the field equations are to be derivable from a least-action principle, and that they would be no higher than the second differential order, involving conventional gravitational and electromagnetic potentials, as well as  a scalar. The following general Lagrangian was proposed 
\begin{equation}
\label{Bergmann}
 L=\abs{g}^{1/2}[f_1(\phi)R+f_2(\phi)M+f_3(\phi)g^{\mu\nu}\phi_{,\mu}\phi_{,\nu}+f_4(\phi)] \,,  
\end{equation}
where $R$ is the (four-dimensional) curvature scalar, and $M$ the Maxwell scalar. This Lagrangian is a scalar density with respect to four-dimensional curvilinear coordinate transformations, and invariant with respect to gauge transformations. The redefinition of the metric tensor field of the kind $g_{\mu\nu}=a(\phi)g_{\mu\nu}^*$, depending on the choice of $a$, makes it possible to replace $f_1$ by a constant (or, if preferred, by $\phi$). Note that, if we choose $f_1=\phi$, the gravity is mediated by one or several long-range scalar fields in addition to the  tensor-scalar theory, different to the action studied by Damour.  

The spontaneous scalarization has been investigated based on action (1) resulting in a wide class of scalar-tensor theories which can pass all the present solar-system tests and still exhibit large,  observable deviations in systems involving neutron stars, providing new motivations for experiments probing the strong-field regime of relativistic gravity, notably binary pulsar experiments \cite{Dcited2}, which might reveal the existence of a scalar contribution to gravity which in solar-system experiments is too small to be detectable. Later, massive scalar fields where introduced \cite{Chen:2015zmx} called dubbed asymmetron. It was showed that the asymmetron undergoes spontaneous scalarization inside dense objects, resulting in possible observational effects. In \cite{Ramazanoglu:2016kul} the effect of a mass term in the spontaneous scalarization of neutron stars was studied, by incorporating the mass term $-m^2_{\phi}\phi^2$ in the action (\ref{actionD}). It was shown spontaneous scalarization can have nonperturbative, strong effects that may lead to observable signatures in binary neutron star or black hole-neutron star mergers, or even in isolated neutron stars. In these models, the evolution of the scalar field is affected by the stress energy tensor of the matter component due to the nonminimal coupling of the scalar field.  Specifically, with $A_{,\phi}=0$ and $A_{,\phi\phi}<0$ at $\phi=\phi_0$, the scalar field in a high density region exhibits tachyonic instability causing the spontaneous scalarization, which leads to interesting phenomenology \cite{Berti:2015itd}.

It is also shown that the Schwarzschild-Newman-Tamburino-Unti solution can get scalarized in models with a nonminimal coupling to either the Gauss-Bonnet or the Chern-Simons terms \cite{Brihaye:2018bgc}. The spontaneous scalarization of the electrically charged black holes in the presence of nonminimal couplings between a scalar field and the Maxwell invariant was also explored, and it was shown through numerical simulations in spherical symmetry that Reissner-Nordstrom black holes evolve into a perturbatively stable scalarized black hole \cite{Herdeiro:2018wub}. In \cite{Kiorpelidi:2023jjw} Scalarization of the Reissner-Nordstr\"om black hole with higher derivative gauge field corrections was studied.  Also, the possibility of spontaneous scalarization of static, spherically symmetric, and asymptotically flat black holes in the Horndeski theory was investigated \cite{Minamitsuji:2019iwp}.   Recently, it was shown that a coupling, with a suitable sign, between a scalar and the GB invariant can lead  to  an  instability  triggered  by  rapid  rotation. Such  instability  is  not  related to superradiance, but is instead tachyonic in nature \cite{Dima:2020yac}. On the other hand, it was study the holographic scalarization of the charged scalar field and it was found that the background black holes are scalarized below a critical temperature \cite{Guo:2020sdu}. Also, spontaneously scalarized black holes have been studied in dynamical Chern-Simons gravity \cite{Doneva:2021dcc} and  
in \cite{Zhang:2022sgt} magnetic-induced spontaneous scalarization in dynamical Chern\textendash{}Simons gravity, and also in Chern-Simons-Kerr theory \cite{Chatzifotis:2022mob}.

Beyond the spontaneous scalarization of black holes, the nonlinear dynamics of scalarized black holes in scalar-Gauss-Bonnet gravity, was studied in \cite{Ripley:2020vpk,Silva:2020omi,Doneva:2021dqn,Kuan:2021lol,East:2021bqk}. In Ref. \cite{Doneva:2021tvn} the authors show the existence of a fully nonlinear dynamical mechanism for the formation of scalarized black holes which is different from the spontaneous scalarization, that is, no tachyonic instability can occur. Although the Schwarzschild black holes are linearly stable against scalar perturbations, for certain choices of the coupling function they are unstable against nonlinear scalar perturbations. This nonlinear instability leads to the formation of new black holes with scalar hair.

Teleparallel Gravity Theories have been extensively studied. These theories have various generalizations such as Teleparallel Supergravity \cite{Salgado:2005pg}, scalar-torsion Gravity Theories \cite{Geng:2011aj, Gonzalez:2014pwa, Kofinas:2015hla, Kofinas:2015zaa, Hohmann:2018vle, Hohmann:2018dqh, Hohmann:2018ijr}, Kaluza-Klein theory for teleparallel gravity  \cite{Geng:2014nfa}, Teleparallel Equivalent of Gauss-Bonnet Gravity \cite{Kofinas:2014owa}, Teleparallel Equivalent of Lovelock Gravity \cite{Gonzalez:2015sha,Gonzalez:2019tky}, Teleparallel Equivalent of higher dimensional gravity theories \cite{Astudillo-Neira:2017anx}, and new classes of modified Teleparallel Gravity models \cite{Bahamonde:2017wwk} were found. Scalarized black hole solutions were also studied in the Teleparallel gravity  by coupling a scalar field with the Gauss-Bonnet invariant  in  \cite{Bahamonde:2022chq}, where  the  scalarization is triggered by the torsion.  Asymptotically flat scalarized black hole solutions were found numerically  and it was shown that, depending on the choice of coupling of the boundary terms, they can have a distinct behaviour compared to the ones known from the usual Einstein Gauss-Bonnet case. More specifically, non-monotonicity of the metric functions and the scalar field can be present, a feature that was not observed until now for static scalarized black hole solutions. Also, previously hairy black hole solutions in three-dimensional spacetime featuring a scalar field coupled to the torsion scalar within teleparallel gravity, along with a potential for the scalar field, were derived in \cite{Gonzalez:2014pwa}.

The aim of this work is to study the effects of the scalarization process  on a TEGR theory. According to GR, curvature is used to geometrize spacetime
and describes the  gravitational interactions.  Then if matter parametrized by a scalar field is coupled to  curvature invariant then new  black holes are obtained in which matter 
is maintained by the interaction with the spacetime curvature. The thermodynamics of these  black holes has been extensively studied in particularly since it bridges the classical theory 
of GR with quantum aspects of gravity and key relationships such as the
Bekenstein-Hawking entropy, which links a black hole event horizon area to its entropy,
and to the laws of black hole thermodynamics. In the TEGR theory the gravitational interaction can be described  in terms of torsion  acting as a force. 
As a consequence, there are no geodesics in the teleparallel equivalent of GR, but only force equations quite analogous
to the Lorentz force equation of electrodynamics.  What we want  to study in this work is
if we consider the simplest GR black hole, the Schwarzschild black hole,  and if we apply the scalarization process using  the teleparallel formalism,
to see if the scalarization of the Schwarzschild black hole is possible and what  are  the thermodynamical properties of the scalarized Schwarzschild black hole 
studying its thermodynamics. 

It would be important to find out if the scalarization procedure  in TEGR gravity theories   can give us interesting physical effects.  Therefore, in the TEGR  gravity theory we consider a scalar field coupled to torsion and we study  the nonlinear scalarization of the Schwarzschild black hole which is triggered by the torsion.  The scalarized black hole solutions are asymptotically flat and studying  their thermodynamics we find that they are meta-stable solutions without phase transitions of first and second order and also we obtain the gravitational energy of the solutions.  We also find that the scalarized black hole is always preferred over the Schwarzschild black hole because its free energy is lower than that of the Schwarzschild black hole. Also, the scalarized black holes are entropically favored over the Schwarzschild solution.

The work is organized as follows: In Sec. \ref{MGT}, we give a brief review of TEGR. Then, in Sec. \ref{Models} we present the model considered and we obtain the equations of motions. Then, we study the scalarization of the  Schwarzschild black hole in Sec. \ref{numeric}. Then, we study the thermodynamic of the scalarized solution in Sec. \ref{Thermo}. Finally, we conclude in Sec. \ref{conclusion}.

\section{Teleparallel Equivalent of General Relativity}
\label{MGT}

The Teleparallel Equivalent of General Relativity  \cite{Unzicker:2005in, Hayashi:1979qx}  corresponds to an equivalent formulation of GR, where the Weitzenb\"{o}ck connection is used to define the covariant derivative instead of the Levi-Civita connection which is used to define the covariant derivative in the context of GR. The Weitzenb\"{o}ck connection has non-null torsion, however it is curvatureless which implies that this formulation of gravity exhibits only torsion. The equivalence between the Riemann and Weitzenb\"{o}ck formulations of GR can be expressed by 
\begin{equation}
\mathcal{T}\, =\, -R + 2\; e^{-1}\;\partial _{\nu }(e\,T_{\sigma }^{\ \sigma \nu
}\,)\,,  \label{divergence}
\end{equation}
where, the left hand side of Eq. (\ref{divergence}) is the Weitzenb\"{o}ck invariant (or torsion scalar), given by 
\begin{equation} \label{scalar-T}
\mathcal{T}\ =\ S_{\rho }^{\ \mu \nu }\ T^{\rho }_{\ \mu \nu }\ =\  \frac{1}{4}T^{\rho \mu \nu}T_{\rho \mu \nu} + \frac{1}{2}T^{\rho \mu \nu}T_{ \nu \mu \rho} - T_{\rho \mu}{}^{\rho}T^{\nu \mu}{}_{\nu}\,,
\end{equation}
where $T^{\rho }_{\ \mu \nu }$ are the components of the torsion two form
$T^{a}=de^{a}$,  in a coordinate basis, coming from the Weitzenb\"{o}ck connection $\Gamma^\lambda_{\nu%
\mu}=\,e_{a}^{\lambda }\,\partial _{\nu }e_{\mu }^{a}$, and  the superpotential $S_{\rho}^{\,\,\, \mu \nu}$ is defined according to
\begin{equation}
S_{\ \mu \nu }^{\rho }=\frac{1}{4}\,(T_{\ \mu \nu }^{\rho }-T_{\mu \nu }^{\
\ \ \rho }+T_{\nu \mu }^{\ \ \ \rho })+\frac{1}{2}\ \delta _{\mu }^{\rho }\
T_{\sigma \nu }^{\ \ \ \sigma }-\frac{1}{2}\ \delta _{\nu }^{\rho }\
T_{\sigma \mu }^{\ \ \,\sigma }.  \label{S}
\end{equation}
In $d$ spacetime dimensions the dynamical fields are the $d$ linearly independent vielbeins and the torsion tensor is formed solely by them and first derivatives of these objects. Actually, $\mathcal{T}$ is the result of a very specific quadratic combination of irreducible representations of the torsion tensor under the Lorentz group $SO(1,3)$ \cite{Hehl:1994ue}. 
In the latter, in TEGR the torsion tensor 
include all the information concerning to the
gravitational field and the action is given by
\begin{eqnarray}  \label{action0}
I = \frac{1}{2 \kappa}\int d^4x e \left(\mathcal{T}+\mathcal{L}_{m}\right),
\end{eqnarray}
where $\kappa =8 \pi G$, $e = \text{det}(e_{\,\, \mu}^a) = \sqrt{-g}$, $\mathcal{T}$ is the torsion scalar and $%
\mathcal{L}_{m}$ stands for the matter Lagrangian. 
The equations of motion can be obtained through the variation of the action (\ref{action0}) with respect to the vierbein, which yields
\begin{eqnarray}\label{eom}
e^{-1}\partial_{\mu}(eS_{a}{}^{\mu\nu})
-e_{a}^{\lambda}T^{\rho}{}_{\mu\lambda}S_{\rho}{}^{\nu\mu}
-\frac{1}{4}e_{a}^{\nu
}\mathcal{T}
= 4\pi Ge_{a}^{\rho}\overset {\mathbf{em}}T_{\rho}{}^{\nu}~,
\end{eqnarray}
where the mixed indixes are used as in $S_a{}^{\mu\nu} =
e_a^{\,\,\, \rho}S_{\rho}{}^{\mu\nu}$. The vierbein field ${\mathbf{e}_a(x^\mu)}$, forms an orthonormal
basis for the tangent space at each point $x^\mu$ of the manifold, that is
$\mathbf{e} _a\cdot\mathbf{e}_b=\eta_{ab}$, with
$\eta_{ab}=diag (1,-1,-1,-1)$. Moreover, the vector $\mathbf{e}_a$ can
be expressed in terms of its components $e_a^{\,\,\, \mu}$ in a coordinate basis,
namely $\mathbf{e}_a=e_a^{\,\,\, \mu}\partial_\mu$. 
Note that the tensor
$\overset{\mathbf{em}}{T%
}_{\rho}{}^{\nu}$ on the right-hand side is the usual energy-momentum
tensor. Therefore, the theory is called ``Teleparallel
Equivalent of General Relativity'' due to the field equations are exactly the same as those of GR for every
geometry choice.

\section{Model of Nonlinear Scalarization}
\label{Models}

In this work, we consider the scalar-torsion model  given by the following action 
\begin{equation}
    I =\int d^{4}x\,e \left[ \left(\frac{1}{2 \kappa}+ f(\phi) \right)\mathcal{T}+\frac{1}{2} g^{\mu \nu} \partial_{\mu} \phi \partial_{\nu}\phi
    \right] \,,
\label{L0}
\end{equation}
where $\kappa=8\pi G$, $\phi$ is the scalar field and $f(\phi)$ is a coupling function of the scalar field with the torsion scalar. It is worth to mentioning that the above action with an $f(\phi) \propto \phi^2$ complemented with an exponential scalar field potential and a matter Lagrangian, was studied in the context of “Teleparallel” Dark Energy  \cite{Geng:2011aj}. Mainly, it was shown that although the minimal case is completely equivalent to standard quintessence, the non-minimal scenario 
exhibit quintessence-like or phantom-like behavior,
or experiencing the phantom-divide crossing.  Then, by using the power-law, the exponential and the inverse hyperbolic cosine
potential ansatzes, it was shown that the model is compatible with observations. In particular, the data favor a nonminimal coupling, and although the scalar field is canonical the
model can describe both the quintessence and phantom regimes \cite{Geng:2011ka}.

In this work we are interested in studying  nonlinear scalarized black hole solutions \cite{Doneva:2021tvn}, \cite{Staykov:2022uwq}, in the scalar torsion model (\ref{L0}); therefore, we consider a coupling function that meets the following conditions
\begin{equation}
\frac{df}{d\phi}(0)=0\,,\quad \frac{d^2 f}{d\phi^2}(0)=0\,.     
\end{equation}
The first condition guarantees the Schwarzschild solution is also a black hole solution to the field equations, for $\phi = 0$, while 
the second condition was imposed to the fact that no tachyonic instability is possible – the Schwarzschild solution is stable against linear scalar perturbations. 
Therefore, as was pointed out in Ref. \cite{Doneva:2021tvn},  if black holes with scalar hair exist, they should form a new black hole phase coexisting with the usual Schwarzschild black hole phase. It will be called a scalarized black hole phase. In the following we consider the case $f(\phi)= \eta\phi^4$, where $\eta$ is a constant which shows the strength of the interaction.

Now, varying the action with respect to the vierbein and the scalar field, the equations of motion respectively are
\begin{eqnarray}\label{eomst}
\notag && \left(\frac{2}{\kappa}+4 f(\phi)\right)\left[ e^{-1}\partial_{\mu}(eS_{a}{}^{\mu\nu})
-e_{a}^{\lambda}T^{\rho}{}_{\mu\lambda}S_{\rho}{}^{\nu\mu}
-\frac{1}{4}e_{a}^{\nu
}\mathcal{T}\right]-\frac{1}{2}e_{a}^{\nu}\partial_{\mu}\phi\partial^{\mu}\phi+4 f'(\phi)(\partial_{\mu}\phi) S_{a}{}^{\mu\nu}\\
&& +e_{a}^{\mu}\partial^{\nu}\phi\partial_{\mu}\phi= 0\,,
\end{eqnarray}
\begin{equation}\label{KGeq}
\frac{1}{e}\partial_\nu(e g^{\mu\nu}\partial_\mu\phi) - f'(\phi) \mathcal{T} = 0  \,.
\end{equation}
Firstly, we assume spherical symmetry. A tetrad that satisfy this condition within the Weitzenb\"{o}ck gauge is \cite{Bahamonde:2022chq} 
\begin{equation}
e^{a}_{\ \mu} =
\begin{pmatrix}
0 & \frac{i}{\sqrt{B(r)}} & 0 & 0\\
i\sqrt{A(r)}\cos\varphi\sin\theta & 0 & -r\sin\varphi & -r\cos\theta\cos\varphi\sin\theta\\
i\sqrt{A(r)}\sin\theta\sin\varphi & 0 & r\cos\varphi  & -r\cos\theta \sin\theta\sin\varphi\\
i\sqrt{A(r)}\cos\theta & 0 & 0 & r\sin^2\theta
\end{pmatrix} \,.
\label{tetrad}
\end{equation}
This complex tetrad also has the same form as the one obtained in \cite{Bahamonde:2021srr} for $f(\mathcal{T},\mathcal{B})$ gravity and \cite{Bahamonde:2022lvh} for a Teleparallel scalar-tensor theory constructed from $\mathcal{T}$, the boundary term $\mathcal{B}$ and the scalar field. Moreover, it was pointed out 
that 
the complex tetrad has simpler equations and exact black hole solutions. This tetrad gives the metric in the standard form in spherical coordinates
\begin{equation}\label{metric1}
ds^2 = A(r)dt^2 - \frac{1}{B(r)}dr^2 - r^2(d\theta^2 + \sin^2\theta d\varphi^2)\,.
\end{equation}
Now, using the tetrad (\ref{tetrad}) the torsion scalar (\ref{scalar-T}) becomes
\begin{equation}
\mathcal{T} = \frac{2 \left(r B(r) A'(r)+A(r) (B(r)+1)\right)}{r^2 A(r)}\,,
    \label{torsionscalar}
\end{equation}
and the tetrad field equations (\ref{eomst}), after contracting them with $e^{a}_{\,\,\, \sigma}$, yield 
\begin{eqnarray}\label{ett}
    E^t_{\;t} & \equiv &  \frac{\left(r B'(r)+B(r)-1\right) \left(2 \eta  \kappa  \phi (r)^4+1\right)}{\kappa  r^2}+\frac{16 \eta  B(r) \phi (r)^3 \phi '(r)}{r}+\frac{1}{2} B(r) \phi '(r)^2\,,\\
\label{err}
    E^r_{\;r} & \equiv & \frac{\left(r B(r) A'(r)+A(r) (B(r)-1)\right) \left(2 \eta  \kappa  \phi (r)^4+1\right)}{\kappa  r^2 A(r)}-\frac{1}{2} B(r) \phi '(r)^2 \,,\\
\label{eang}
\notag    E^\theta_{\;\theta} = E^\varphi_{\;\varphi} & \equiv & \Big( -r B(r) A'(r)^2 \left(2 \eta  \kappa  \phi (r)^4+1\right)+A(r) \big(2 r B(r) A''(r) \left(2 \eta  \kappa  \phi (r)^4+1\right)\\
&&
 + A'(r) \left(\left(r B'(r)+2 B(r)\right) \left(2 \eta  \kappa  \phi (r)^4+1\right)+16 \eta  \kappa  r B(r) \phi (r)^3 \phi '(r)\right)\big)\\
&& \notag + 2 A(r)^2 \left(B'(r) \left(2 \eta  \kappa  \phi (r)^4+1\right)+\kappa  B(r) \phi '(r) \left(16 \eta  \phi (r)^3+r \phi '(r)\right) \right) \Big) / (4 \kappa  r A(r)^2)\,,
\end{eqnarray}
while the modified Klein-Gordon equation (\ref{KGeq}) becomes
\begin{eqnarray}\label{ekg}
 \notag E_{\phi} & \equiv &  -\frac{\phi '(r) \left(r B(r) A'(r)+A(r) \left(r B'(r)+4 B(r)\right)\right)}{2 r A(r)}-\frac{8 \eta  \phi (r)^3 \left(r B(r) A'(r)+A(r) (B(r)+1)\right)}{r^2 A(r)}\\
&& -B(r) \phi ''(r)\,. 
\end{eqnarray}

It is worth mentioning that
when the scalar field is constant  ($\phi=\phi_0$), the theory admits the Schwarzschild solution, i.e.
\begin{equation}
    A(r)=B(r)=1 -\frac{2 G M}{r}\,,
\end{equation}
where $M$ is the black hole mass, and the radius of the horizon is located at $r_H=2 G M$. 

\begin{figure}[ht]
\begin{center}
\includegraphics[width=0.4\textwidth]{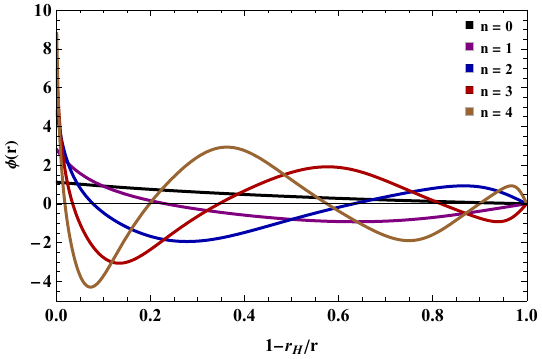}
\includegraphics[width=0.4\textwidth]{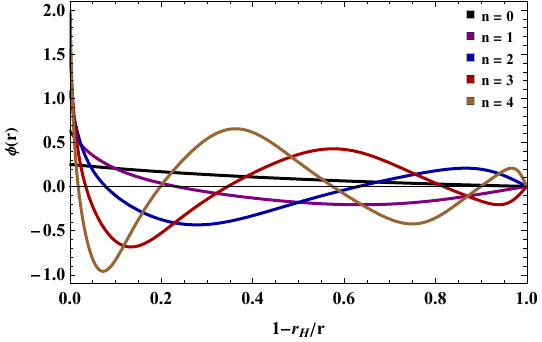}
\end{center}
\caption{Different branches of the scalar field in the decoupling limit labelled by the number of nodes $n$, for $\eta=0.1$ (left panel) and $\eta=2$ (right panel).}
\label{scalar}
\end{figure}

Now we investigate how the scalarized branches of the Schwarzschild black hole can be exist, for that we need to consider the decoupling limit approximation, that is, we consider the nonlinear Klein-Gordon equation (\ref{KGeq}) and the geometry kept fixed, given by the Schwarzschild solution. Considering a static spherically symmetric scalar field $\phi = \phi(r)$ we get
\begin{equation}
   16 \eta  \phi (r)^3+r (r-r_H) \phi ''(r)+(2r-r_H) \phi '(r)=0\,,
\end{equation}
to solve this equation, we use the shooting method \cite{stoer1980introduction, press2007numerical}. For this, it is necessary to consider the Taylor expansions of the scalar field near the event horizon, $r_H$
\begin{equation}
    \phi(r) = \phi_H + \phi_H (1-16\eta \phi_H^2)(r-r_H) + \phi_H (1-24\eta \phi_H^2+192 \eta ^2\phi_H^4)(r-r_H)^2 +...\,,
\end{equation}
and near spacial infinity ($r\rightarrow \infty$)
\begin{equation}
    \phi(r) = \phi_\infty + \frac{\phi_\infty}{2r} + \frac{\phi_\infty(1-8\eta \phi_\infty^2)}{3r^2} + \frac{\phi_\infty(1-16\eta \phi_\infty^2)}{4r^3} + ...\,.
\end{equation}
The solution is determined by three parameters, $\{\phi_H,\phi_\infty,\eta\}$. When the value of $\eta$ is fixed, specific values of $\phi_H$ and $\phi_\infty$ allow for a bounded solution. In Fig. (\ref{scalar}), we show the scalar field profile for the decoupling limit approximation, where, for fixed values of $\eta = \{0.1,2\}$, distinct branches of solutions can coexist, classified by $n$ the number of nodes.  

\section{Scalarized Black Hole Solutions}
\label{numeric}

 In this section we present the scalarized black holes solutions obtained by solving the full system of equations (\ref{ett})-(\ref{ekg}) numerically. To facilitate obtaining numerical solutions, we introduce a bounded radial coordinate $z=1-r_H/r$, and redefine the metric functions and the scalar field as
\begin{equation}\label{IRA}
    A(z) = z a(z)\,,\quad B(z)=zb(z)\,, \quad \phi(z)=(1-z)\psi(z)\,.
\end{equation}

Now, by considering the Taylor expansions of the metric functions and the scalar field near the event horizon ($z=0$), we have the following:
\begin{eqnarray}
\nonumber a(z) &=& a_H + a_{H1}(a_H,\psi_H,\eta) z + a_{H2}(a_H,\psi_H,\eta)z^2 +...\,, \\
b(z) &=& b_H + b_{H1}(b_H,\psi_H,\eta) z + b_{H2}(b_H,\psi_H,\eta)z^2 +...\,, \\
\nonumber \psi(z) &=& \psi_H + \psi_{H1}(\psi_H,\eta) z + \psi_{H2}(\psi_H,\eta)z^2 +...\,, 
\end{eqnarray}
and then substituting them into the field equations (\ref{ett})-(\ref{ekg}), we obtain

\begin{eqnarray}\label{IRB}
\nonumber    a(z) &=& a_H + \frac{64 a_H \eta ^2 \kappa  \psi_H^6} {2 \eta  \kappa \psi_H^4+1} z + \frac{64 a_H \eta ^2 \kappa  \psi_H^6 \left(96 \eta ^2 \kappa  \psi_H^6+2 \eta  \kappa  \psi_H^4-48 \eta  \psi_H^2+1\right)}{3 \left(2 \eta  \kappa  \psi_H^4+1\right)^2} z^2 +...\,,\\
\nonumber    b(z) &=& 1 + \frac{64 \eta ^2 \kappa  \psi_H^6}{2 \eta  \kappa  \psi_H^4+1} z + \frac{64 \eta ^2 \kappa  \psi_H^6 \left(2 \eta  \kappa  \psi_H^4-192 \eta  \psi_H^2+1\right)}{3 \left(2 \eta  \kappa  \psi_H^4+1\right)^2} z^2 +...\,,\\
\psi(z) &=& \psi_H + \psi_H(1-16 \eta \psi_H^2) z \\
&& + \frac{\psi_H \left(640 \eta ^3 \kappa  \psi_H^8-48 \eta ^2 \kappa  \phi_H^6+192 \eta ^2 \psi_H^4+2 \eta  \kappa  \psi_H^4-24 \eta  \psi_H^2+1\right)}{2 \eta  \kappa  \psi_H^4+1} z^2 +...\,.
\end{eqnarray}
Then, by applying the same procedure at infinity, that is,  by considering the Taylor expansions of the metric functions and the scalar field near spatial infinity ($z \rightarrow 1$), we have the following:
\begin{eqnarray}
\nonumber  a(z) &=& a_\infty + a_{1}(a_0,b_1) (1-z) +a_{2}(a_0,b_1)(1-z)^2 
+...\,, \\
b(z) &=& b_\infty + b_{1} (1-z) + b_{2}(b_0,b_1,\psi_\infty)(1-z)^2 
+...\,, \\
\nonumber  \psi(z) &=& \psi_{\infty} + \psi_{1}(b_1,\psi_\infty)(1-z) + \psi_{2}(b_1,\psi_\infty,\eta)(1-z)^2 + ...\,,
\end{eqnarray}
and then substituting them into the field equations (\ref{ett})-(\ref{ekg}), we obtain
\begin{eqnarray}\label{IRB}
 \nonumber   a(z) &=& a_{\infty} + a_1(1-z) -  a_1(1-z)^2+ \frac{1}{12}(12a_1+\kappa a_{\infty}\psi_{\infty}^2-\kappa a_1 \psi_{\infty}^2)(1-z)^3 + ...\,,\\
 \nonumber   b(z) &=& b_{\infty} + \frac{a_1}{a_{\infty}}(1-z) +
\frac{2a_1+\kappa a_{\infty} \psi_{\infty}^2}{2 a_{\infty}}(1-z)^2 + \frac{4a_1+3\kappa a_{\infty}\psi_{\infty}^2-\kappa a_1 \psi_{\infty}^2}{4a_{\infty}} (1-z)^3 + ...\,,\\
\psi(z) &=& \psi_{\infty} + \frac{\psi_{\infty}(a_{\infty}-a_1)}{2a_{\infty}} (1-z) \\
&& + \frac{\psi_{\infty}(4a_{\infty}^2-8a_{\infty}a_1+4a_1^2-32\eta a_{\infty}^2\psi_{\infty}^2-\kappa a_{\infty} ^2\psi_{\infty}^2)}{12a_{\infty}^2}(1-z)^2 +...\,.
\end{eqnarray}

As we are searching for asymptotically flat scalarized black holes, we claim the  boundary conditions $a_{\infty}= 1$, $b_{\infty} = 1$,  where $a_{\infty}$ and $b_{\infty}$ represent the asymptotic behaviour of the functions $a(z)$ and $b(z)$, respectively, and $\phi(z \rightarrow 1) \rightarrow 0$, this last, in view of Eq. (\ref{IRA}), is satisfied requiring $\psi(z)$ to be regular at $z=1$. The parameter $b_1$,  which represents the coefficient of $(1-z)$ in the expansion of $b(z)$ at infinity, is related to the ADM mass of the scalarized black hole solutions (see next section). The motion equations also set $b_1=a_1$ and $b_H \equiv b(z=0)=1$. The solution is determined by a set of five parameters $\{\eta, a_H, \psi_H, \psi_\infty, b_1\}$.  When the value of $\eta$ is fixed, specific values of $\{a_H, \psi_H,\psi_\infty, b_1\}$ allow for a bounded solution where $\phi(z \rightarrow 1) \rightarrow 0$ to be achieved.  The differential equations are solved using the shooting method. Specifically, we perform a match between the solutions at a specific value of \( z \), namely \( z_{\text{match}} = 0.5 \), with a precision of \( \epsilon = 10^{-8}\) and \( \delta = 10^{-7} \). Here, \( \epsilon \) represents the tolerance for the agreement between the solutions from the horizon and infinity, ensuring that they match with a difference smaller than \( 10^{-5} \) at \( z_{\text{match}} \). The parameter \( \delta \), on the other hand, represents the tolerance for the solution parameters.  
In each expansion of the metric functions and the scalar field, we include 10 terms in the expansion, both at the horizon and at infinity. Actually, for fixed $\eta$, there exists an countable set of the coupling constant, $\{({a_H, \psi_H,\psi_\infty, b_1};n)\}_{n=0}^{n_{max}}$, which can support the bounded scalar with $n$ labeling the number of nodes of the solution, and $n_{max}$ is the maximum number of nodes for that value of $\eta$.  Some solutions are shown in Table \ref{parametersolutionsA} of Appendix \ref{Solutions} . However,  we will focus on the solutions presented in Table \ref{parametersolutions} in order to study their behaviour and thermodynamics. Notice that for small values of $\eta$ only solutions without nodes exist, while for higher values of $\eta$ solutions with more nodes appear until there are different solutions with the same number of nodes.


\begin{table}[ht]
  \centering
  \scalebox{0.9} {
  \begin{tabular}{|c|c|c|c|c|c|c|c|}
    \hline
    \textbf{Type of BH} & $\boldsymbol{\eta}$ & \textbf{Number of nodes} & $\boldsymbol{a_H}$ & $\boldsymbol{\psi_H}$ & $\boldsymbol{\psi_\infty}$ & $\boldsymbol{b_1}$ & \textbf{Label}\\
    \hline
    Schwarzschild & 0 & {} & 1 & 0 & 0 & 0 & {Schw}\\
    \hline
      \multirow{2}{*}{Scalarized} & \multirow{2}{*}{0.065}  & 0 & $1.153 \times 10^{-6}$ & 9.672 & 0.418 & -0.222 & Ia\\
    \cline{3-8}
    &  & 0 & 0.521 & 1.468 & 0.663 & -0.159 & Ib\\
    \hline
    \multirow{3}{*}{Scalarized} &   \multirow{3}{*}{2} & 0 & 0.979 & 0.250 & 0.125 & -0.005 & IIa\\
    \cline{3-8}
    &  & 1 & 0.004 & 1.410 & -0.540 & -0.070 & IIb\\
    \cline{3-8}
    &  & 1 & 0.205 & 0.764 & -0.719 & -0.068 & IIc\\
    \hline
    \multirow{4}{*}{Scalarized} &  \multirow{4}{*}{20} & 0 & 0.998 & 0.079 & 0.040 & -0.0005 & IIIa\\
    \cline{3-8}
    &  & 1 & 0.890 & 0.203 & -0.265 & -0.007 & IIIb\\
    \cline{3-8}
    &  & 2 & 0.026 & 0.607 & 0.649 & -0.032 & IIIc\\
    \cline{3-8}
    &  & 2 & 0.108 & 0.481 & 0.704 & -0.032 & IIId\\
    \hline
  \end{tabular}}
  \caption{Values of the parameters $\eta$, $n$, $a_H$, $\psi_H$, $\psi_\infty$, and $b_1$ for the Schwarzschild and scalarized black hole solutions for $\eta=0.065,2,20$.}
  \label{parametersolutions}
\end{table}

\newpage

\begin{figure}[ht]
\begin{center}
\includegraphics[width=0.4\textwidth]{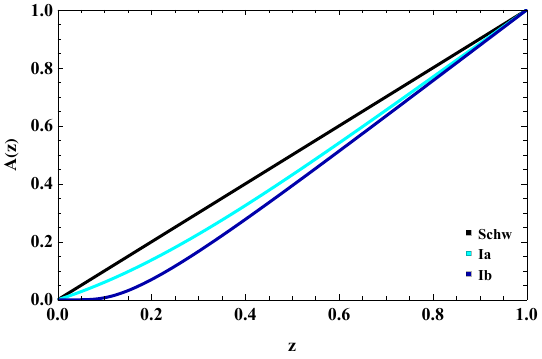}
\includegraphics[width=0.4\textwidth]{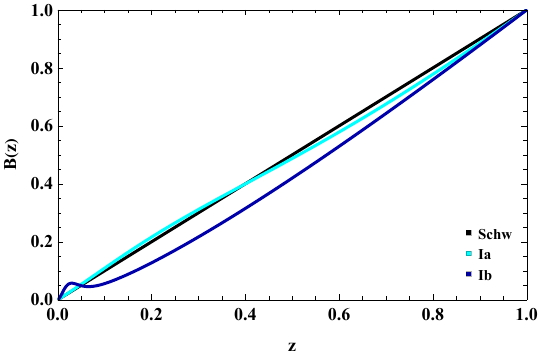}\\
\includegraphics[width=0.4\textwidth]{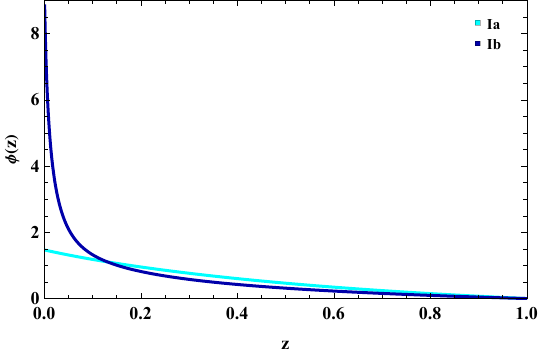}
\includegraphics[width=0.4\textwidth]{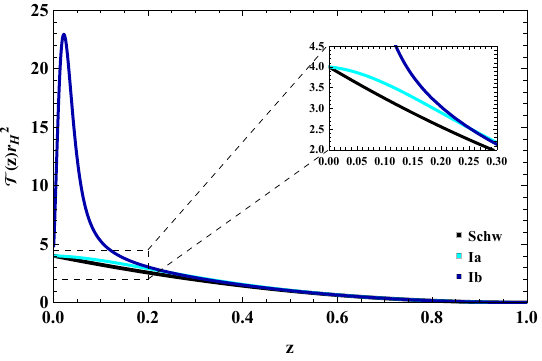}
\end{center}
\caption{The behaviour of $A(z)$ (top left panel), $B(z)$ (top right panel), $\phi(z)$ (bottom left panel), and $\mathcal{T}(z)r_H^2$ (bottom right panel) as a function of $z$. Here $\eta=0.065$.}
\label{ABPhi0}
\end{figure}

\begin{figure}[ht]
\begin{center}
\includegraphics[width=0.4\textwidth]{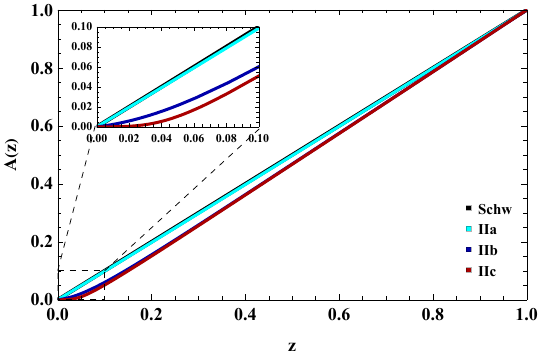}
\includegraphics[width=0.4\textwidth]{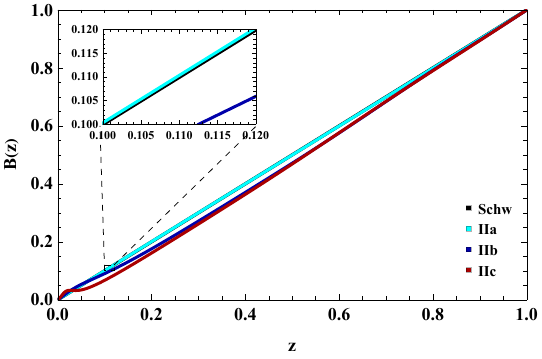}\\
\includegraphics[width=0.4\textwidth]{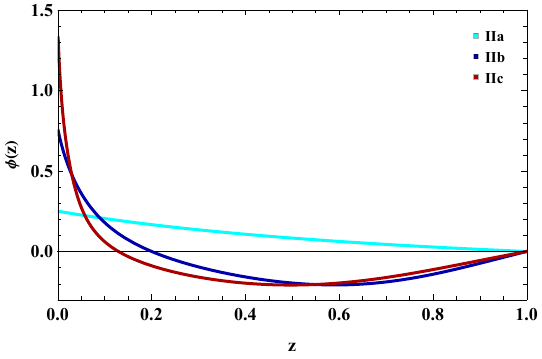}
\includegraphics[width=0.4\textwidth]{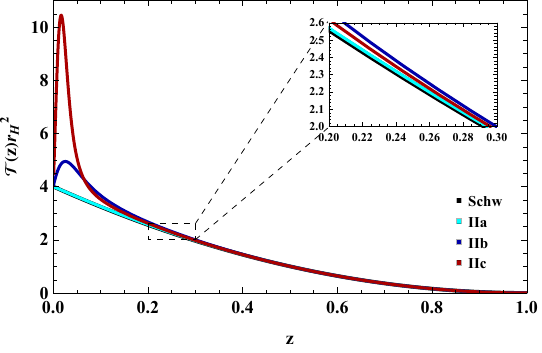}
\end{center}
\caption{The behaviour of $A(z)$ (top left panel), $B(z)$ (top right panel), $\phi(z)$ (bottom left panel), and $\mathcal{T}(z)r_H^2$ (bottom right panel) as a function of $z$.  Here, $\eta=2$.}
\label{ABPhi1}
\end{figure}

\begin{figure}[ht]
\begin{center}
\includegraphics[width=0.4\textwidth]{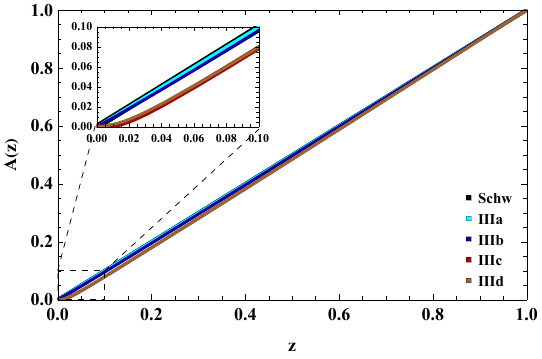}
\includegraphics[width=0.4\textwidth]{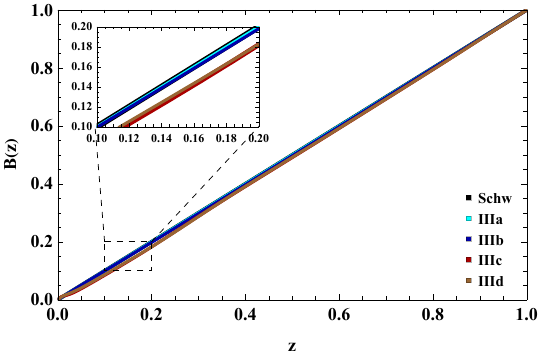}\\
\includegraphics[width=0.4\textwidth]{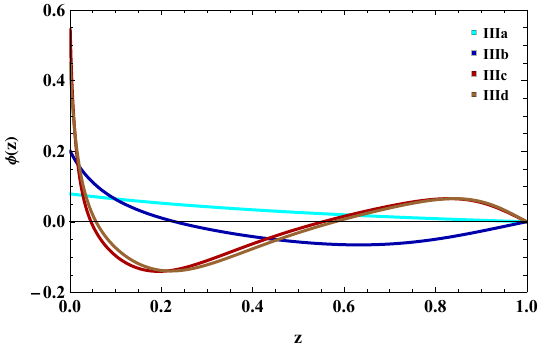}
\includegraphics[width=0.4\textwidth]{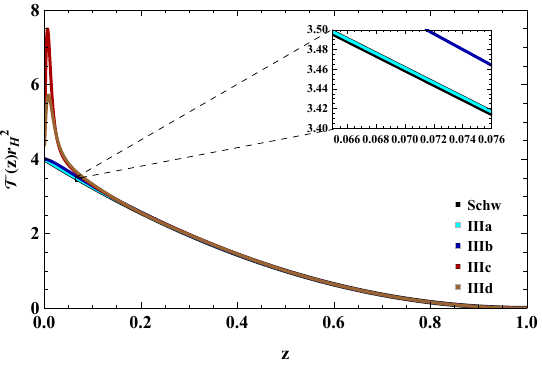}
\end{center}
\caption{The behaviour of $A(z)$ (top left panel), $B(z)$ (top right panel), $\phi(z)$ (bottom left panel), and $\mathcal{T}(z)r_H^2$ (bottom right panel) as a function of $z$. Here, $\eta=20$. }
\label{ABphi2}
\end{figure}

In the following figures we show the behaviour of $A(z)$, $B(z)$, $\phi(z)$, and $\mathcal{T}(z)r_H^2$ for $\eta=0.065$ (Fig. \ref{ABPhi0}), $\eta=2$ (Fig. \ref{ABPhi1}), and $\eta=20$ (Fig. \ref{ABphi2}). In these figures we have plotted the Schwarzschild case too, in order to shown the deviations from the Schwarzschild case, for the scalarized solutions.  For $\eta=0.065$, there are two solutions with zero nodes, labeled as Ia and Ib (see Table \ref{parametersolutions}). Solution Ia corresponds to small deviations from the Schwarzschild case, except for the scalar field, while solution Ib  corresponds to larger deviations from the Schwarzschild case, as shown in the metric functions $A(z)$, $B(z)$, and the torsion scalar in Fig. \ref{ABPhi0}. Note that the metric functions are positive definite outside the event horizon. Concerning to the scalar field, solution Ib presents a scalar field at the event horizon that is greater than the one for solution Ia. We observe that the numerical solutions for the scalar field are regular and satisfy the boundary condition at infinity, i.e., $\phi\rightarrow 0$ when $z \rightarrow 1$. Additionally, the behaviour of the torsion scalar for solution Ia presents a maximum value at the event horizon, while for solution Ib, the torsion scalar increases to a maximum value greater than that for solution Ia, very close to the event horizon, but not exactly at the event horizon, and then starts to decrease. Note that the torsion scalar $\mathcal{T}$ is positive and regular at the event horizon ($z=0$), and it vanishes at infinity $(z \rightarrow 1)$.


\newpage

For $\eta=2$, there is one solution with zero node (IIa), and two solutions with one node (IIb, IIc). Solution IIa corresponds to small deviations from the Schwarzschild case, except for the scalar field, similar to the small deviations observed for $\eta=0.065$. However, solutions IIb and IIc correspond to larger deviations from the Schwarzschild case, as shown in the metric functions $A(z)$, $B(z)$, and the torsion scalar in Fig. \ref{ABPhi1}, with the metric functions being positive definite outside the event horizon. Concerning to the scalar field, solution IIc presents a node at a smaller $z$-value than solution IIb, and the scalar field at event horizon is greater than in solution IIb. We observe that the numerical solutions for the scalar field are regular and satisfy the boundary condition at infinity, i.e., $\phi\rightarrow 0$ as $z \rightarrow 1$. Additionally, for solutions IIb and IIc the torsion scalar increases to a maximum value, greater than the maximum for solution IIa. This maximum occurs very close to the event horizon, but not exactly at it, and then the torsion scalar begins to decrease. The maximum value of the torsion scalar for solution IIc is greater than for solution IIb, with the torsion scalar $\mathcal{T}$ being positive and regular at the event horizon ($z=0$) and vanishing at infinity $(z \rightarrow 1)$.

\newpage

For $\eta=20$, there is one solution with zero node (IIIa), one solution with one node (IIIb,) and two solutions with two nodes (IIIc, IIId). In this case, solutions IIIa and IIIb correspond to small deviations from the Schwarzschild case, except for the scalar field, similar to the small deviations from the Schwarzschild case for $\eta=0.065$. 
However, solutions IIIc and IIId correspond to larger deviations from the Schwarzschild case, 
as shown in the torsion scalar in Fig. \ref{ABphi2}. The metric functions are positive definite outside the event horizon. Concerning to the scalar field, solution IIIc has its nodes at smaller $z$-values than solution IIId, and the scalar field at the event horizon is greater than for solutions IIIa and IIIb. We can observe that the numerical solutions for the scalar field are regular and satisfy the boundary condition at infinity, i.e., $\phi\rightarrow 0$ as $z \rightarrow 1$. Additionally, for solutions IIIc and IIId, the torsion scalar increases to a maximum value greater than the maximum for solutions IIIa and IIIb. This maximum occurs close to, but not exactly at, the event horizon, and then the torsion scalar starts to decrease. The maximum value of the torsion scalar for solution IIIc is greater than that for solution IIIb, with the torsion scalar $\mathcal{T}$ being positive and regular at the event horizon ($z=0$) and vanishing at infinity $(z=1)$.

\newpage

It is worth mentioning that, for the cases analyzed and values of $\eta$ that present more than one solution with the same number of nodes, the maximum value of the torsion scalar corresponds to the solution with nodes closest to the event horizon.

\section{Thermodynamics}
\label{Thermo}

In this section first we will construct an energy-momentum pseudo-current from which it is possible to obtain the gravitational energy of the solutions. Then, we apply the method developed by Padmanabhan \cite{Padmanabhan:2012gx}, in order to determine the entropy, by rewriting the equations of motion as the first law of thermodynamics. Finally, we employ the Wald's formalism to derive the conserved quantities and the conservation laws. It is worth mentioning that the thermodynamics quantities in Teleparallel Gravity theories depend on the choose of the tetrads \cite{Fiorini:2023axr}.

\subsection{Energy-momentum pseduo-current}

In order to find the energy of the gravitational field we consider the energy-momentum pseudo current. It is worth to mention that similar pseudo-current where constructed in the context of TEGR \cite{Bras,Maluf1,Maluf2}. So, by considering that the field equations (\ref{eomst}) can be written as
\begin{eqnarray}\label{ne}
\notag  e^{-1} \partial_{\mu} \left(\left(\frac{2}{\kappa}+4 f(\phi) \right) eS_{a}{}^{\mu\nu} \right) &=&
\left(\frac{2}{\kappa}+4 f(\phi) \right) \left(e_{a}^{\lambda}T^{\rho}{}_{\mu\lambda}S_{\rho}{}^{\nu\mu}
+\frac{1}{4}e_{a}^{\nu
}\mathcal{T}  \right)+\frac{1}{2}e_{a}^{\nu}\partial_{\mu}\phi\partial^{\mu}\phi \\
&& -e_{a}^{\mu}\partial^{\nu}\phi\partial_{\mu}\phi\ \,,
\end{eqnarray}
and defining
\begin{equation}
j_a^{\,\,\nu}=\left(\frac{2}{\kappa}+4 f(\phi) \right) \left(e_{a}^{\lambda}T^{\rho}{}_{\mu\lambda}S_{\rho}{}^{\nu\mu}
+\frac{1}{4}e_{a}^{\nu
}\mathcal{T}  \right) \,,
\label{corriente}
\end{equation}
the motion equations (\ref{ne}) can be cast into the form
\begin{equation} \label{eqofm}
e^{-1} \partial_{\mu} \left(  \left(\frac{2}{\kappa}+4f(\phi) \right) eS_{a}{}^{\mu\nu} \right) = 
j_a^{\,\, \nu}+ \mathfrak{T}_a^{\,\, \nu} \,,
\end{equation}
where
\begin{equation}
\mathfrak{T}_a^{\,\, \nu} = \frac{1}{2}e_{a}^{\nu}\partial_{\mu}\phi\partial^{\mu}\phi
 - e_{a}^{\mu}\partial^{\nu}\phi\partial_{\mu}\phi \,,
\end{equation}
stands for the energy-momentum tensor of the scalar field. In all these expressions the \emph{superpotential} $S^a_{\,\,\, \mu \nu} = e^a_{\,\,\, \alpha} S^{\alpha}_{\,\,\, \mu \nu}$ plays a prominent role and verifies $\mathcal{T}=S^{a\mu}_{\,\,\,\,\,\,\lambda}T_{a\mu}^{\,\,\,\,\,\,\lambda}$. Due to the skew-symmetry of $S^{a}_{\,\,\, \mu \nu}$ with respect to the last two indexes, a natural conserved current appears
\begin{equation}
\partial ^{\nu}[e \,J^a_{\,\,\nu}]=0\,,\,\,\,\,\,\,J^a_{\,\,\nu}=j^a_{\,\,\nu}+ \mathfrak{T}^a_{\,\,\nu}\,.
\end{equation}
In this way, $j^a_{\,\,\nu}$ is viewed as the energy-momentum current of the gravitational field. Because of that, we can define the total energy-momentum four-vector of the gravitational field by integrating the $\nu=t$ component of $J^a_{\,\,\nu}$ over a spatial hypersurface $V$ defined by $t=$constant (we shall use $J^{a\nu}$ instead of  $J^a_{\,\,\nu}$), namely 
\begin{equation}
p^a= \int _V d^3x \,e \,J^{a t}\,.
\label{vem}
\end{equation}
So, using the motion equations, Eq. (\ref{eqofm}) can be written as:
\begin{equation}
p^a=  \int _V d^3x \, \partial_{\mu} \left(\left(\frac{2}{\kappa}+4 f(\phi) \right) e S^{a \mu t} \right)=  \oint _{S(V)} dS_{\mu}\, \left(\frac{2}{\kappa}+4f(\phi) \right) e S^{a \mu t}  \,.
\label{vem2}
\end{equation}
Therefore, the total energy present in the gravitational and matter fields contained in the volume $V$ is just
\begin{equation}
\mathcal{E}=p^{(0)}=  \oint _{S(V)} dS_{\mu}\, \left(\frac{2}{\kappa}+4 f(\phi) \right) e S^{(0) \mu t} \,.
\label{vem3}
\end{equation}
Finally, due to the geometry under consideration, we are interested in the components $S^{(0)r t}$; remember that the normal vector is radial, so $\mu=r$. We find
\begin{equation}\label{endenfin}
  S^{(0) r t}= 0 \,.
\end{equation}
In this way, we see that the total gravitational energy is just $\mathcal{E}= 0$. \\

\subsection{Padmanabhan method}

Now, in order to determine the black hole entropy we evaluate the equation of motion (\ref{err}) at the horizon, which yields 
\begin{equation}
(b_H-1)\left(\frac{1}{\kappa}+2 \eta\psi^4_H\right)=0\,,
\end{equation}
which can be written as
\begin{equation}
\label{Padma}
    \frac{\sqrt{a_H b_H}}{4\pi r_H}d\left[8\pi^2\left(\frac{1}{\kappa}+2\eta \psi_H^4\right)r_H^2\right]=d\left[4\pi\sqrt{\frac{a_H}{b_H}}\left(\frac{1}{\kappa}+2\eta\psi^4_H\right)r_H\right]\,.
\end{equation}
 The Hawking temperature is given by $T = \frac{\tilde{\kappa}}{ 2 \pi }$, where $\tilde{\kappa}$ is the surface gravity, defined by $\xi^{\mu} \nabla_{\mu} \xi_{\nu} = \tilde{\kappa} \xi_{\nu}$, where $\xi^{\mu} = (1,0,0,0)$ is the timelike Killing vector field. Here, $\nabla_{\mu}$ is the Levi-Civita covariant derivative. So, the Hawking temperature yields
\begin{equation}
    T=\frac{\sqrt{a_H b_H}}{4\pi r_H}\,.
\end{equation}
Therefore, Eq. (\ref{Padma}) is manifestly written in the form of the first law of black hole thermodynamics as $ Td\mathcal{S}=d\mathcal{M}$. Then, it is natural to identify the scalarized black hole mass $\mathcal{M}$ as
\begin{equation}
    \mathcal{M}=4\pi r_H\sqrt{\frac{a_H}{b_H}} \left(\frac{1}{\kappa}+2\eta\psi_H^4\right)=\frac{a_H}{T}\left(\frac{1}{\kappa}+2\eta \psi_H^4\right)\,,
\end{equation}
and the scalarized black hole entropy $\mathcal{S}$ as
\begin{equation} \label{Entropy}
    \mathcal{S}=8\pi^2 r_H^2\left(\frac{1}{\kappa}+2\eta\psi_H^4\right)=\frac{A}{4 G} + 4\pi f(\psi_H)A\,,
\end{equation}
where $A= 4 \pi r_H^2$ is the area of the black hole surface. Note that the entropy, for $f(\psi_H)=0$ or $\phi=0$, reduces to the Bekenstein-Hawking entropy $\mathcal{S} = \frac{A}{4 G}$.

\subsection{Noether charge}

In order to obtain the conserved charges and conservation laws that follow from the diffeomeorfism invariance for the scalarized asymptotically flat black hole solutions obtained previously, we use the Wald's formalism \cite{Wald:1993nt, Iyer:1994ys, Wald:1999wa} developed in the context of GR, and we extend it for Teleparallel Gravity with a scalar field non-minimally coupled with the torsion scalar. The Noether charge in TEGR was studied in \cite{Hammad:2019oyb}, also in \cite{Emtsova:2019moq} expressions for conserved currents were constructed applying the Noether theorem. Additional references discussing applications of Noether charges include \cite{Herrera:2017ztd, Dengiz:2020fpe, Anastasiou:2021tlv}. So, in the next, to simplify the notation, we denote the dynamical variables as $\chi = \{ e^a_{\,\, \mu}, \phi \}$. Thus, the variation of the  Lagrangian with respect to the dynamical fields of the theory can be written as
\begin{equation}
\delta L = E_{\chi}[\chi] \delta \chi + \partial_{\mu} \left(  e \Theta^{\mu}[\chi, \delta \chi] \right)\,,
\end{equation}
where
\begin{equation} \label{BT}
\Theta^{\mu} = 4 \left( \frac{1}{2 \kappa} + f(\phi) \right) S_{a}^{\,\,\, \mu \nu} \delta e^a_{\,\, \nu} + g^{\mu \nu} \partial_{\nu} \phi \delta \phi \,.
\end{equation}
Denoting the generator of the diffeomorphism as $\xi^{\mu}$, the Noether charge associated to it can be obtained from this potential, by replacing the general variations $\delta e^a_{\,\, \nu}$ and $\delta \phi$ by the Lie derivatives $\mathcal{L}_{\xi} e^a_{\,\, \nu}$ and $\mathcal{L}_{\xi} \phi$ repectively, induced by the diffeomorphism $\xi^{\mu}$. Then, it can be defined the on-shell Noether current when the equations of motion are satisfied $E_{\chi} = 0$ as follows
\begin{equation} \label{current}
J^{\mu}[\chi] = e \Theta^{\mu}[\chi, \mathcal{L}_{\xi} \chi] - \xi ^{\mu} L[\chi] \,,
\end{equation}
where
\begin{equation}
L = e \left[ \left(\frac{1}{2 \kappa}+f(\phi) \right)\mathcal{T}+\frac{1}{2} g^{\mu \nu} \partial_{\mu} \phi \partial_{\nu}\phi
    \right]\,,
\end{equation}
and it is conserved on-shell $\partial_{\mu} J^{\mu} \simeq 0$, here the symbol $\simeq$ means on-shell equality. Furthermore, using the Poincar\'e lemma, we can express the Noether current locally as $J^{\mu}[\chi] \simeq  \partial_{\nu} ( e Q^{\mu \nu}$), where $Q^{\mu \nu}$ is known as the Noether prepotential. This prepotential can be constructed replacing in Eq. (\ref{current}), the expressions for $\Theta^{\mu}$ and $L$, using $\mathcal{L}_{\xi} e^a_{\,\, \nu} = \xi^{\rho} \partial_{\rho} e^a_{\,\, \nu} + e^a_{\,\, \rho} \partial_{\nu} \xi^{\rho}$ and $\mathcal{L}_{\xi} \phi = \xi ^{\alpha}  \partial_{\alpha} \phi$, and the field equations, we arrive at
\begin{equation}
Q^{\mu \nu} = - 4 \left( \frac{1}{2 \kappa}+ f(\phi)  \right) S_a^{\,\, \mu \nu} \xi^a \,.
\end{equation}
This charge yield the entropy of the black hole, when the generator of the diffeomorphism is a Killing vector field. Then, a Hamiltonian can be defined by
\begin{equation}
H[\xi] = \int_{\Sigma} d \Sigma_{\mu \nu} \left( Q^{\mu \nu} - \xi^{[\mu}B^{\nu ]} \right)\,, 
\end{equation}
where $d \Sigma_{\mu \nu}$ is the volume element of the codimension-2 surface $\Sigma$ with induced metric $h_{\mu \nu}$ and can be expressed as $d \Sigma_{\mu \nu} = \frac{1}{2}\frac{1}{2!} \sqrt{h} \epsilon_{\mu \nu \rho \sigma} dx^{\rho} \wedge dx^{\sigma}$, where $h = det (h_{\mu \nu})$. The Hamiltonian describe the dynamics generated by the vector field $\xi^{\mu}$ when the boundary term $B$ exists. For this, the following condition must be satisfied \cite{Iyer:1994ys}
\begin{equation} \label{cond1}
\int_{\Sigma} d\Sigma_{\mu \nu} \xi^{[\mu} \Theta^{\nu]}[\chi, \delta \chi] = \delta \int_{\Sigma} d\Sigma_{\mu \nu} \xi^{[ \mu}B^{\nu ]}[\chi] \,,
\end{equation}
and $\delta \chi$ satisfies the linear equations of motion.

For the generator of the asymptotic time translation $\xi = \partial_t$, the gravitational energy can be defined by
\begin{equation}
\mathcal{E} = \int_{\infty} d\Sigma_{\mu \nu} \left( Q^{\mu \nu} - t^{[\mu}B^{\nu]} \right) \,.
\end{equation}
For the scalarized asymptotically flat solutions, the superpotential $S_a^{\,\, \mu \nu} \sim \mathcal{O}(r^{-1})$ and $\phi \sim \mathcal{O}(r^{-1})$ asymptotically; therefore, using (\ref{BT}) we obtain
\begin{equation}
\int_{\infty} d\Sigma_{\mu \nu} \xi^{[\mu} \Theta^{\nu]}[\chi, \delta \chi] = \int_{\infty} d \Sigma_{tr} \xi^t \Theta^r \,,
\end{equation}
where
\begin{equation}
\Theta^r = \frac{2}{\kappa} S_a^{\,\,\,r \mu} \delta e^a_{\,\,\, \mu}+ \mathcal{O}(r^{-3}) = \delta \left( - \frac{r_H (1-b_1)}{\kappa r^2}\right) + \mathcal{O}(r^{-3}) \,.
\end{equation}
Therefore, the contribution of the boundary term $B$ to the gravitational energy  is given by $\int_{\infty} d \Sigma_{tr} \xi^t B^r = - 4 \pi r_H (1-b_1)/\kappa$. On the other hand, the contribution of the term $Q$ to the energy is null due to $S^{(0)rt}=0$.
So, the gravitational energy can be expressed as
\begin{equation} \label{mass}
\mathcal{E} = \frac{4\pi r_H (1-b_1)}{\kappa} = \frac{\sqrt{a_H}(1-b_1)}{\kappa T}  \,,
\end{equation}
this expression corresponds to the ADM mass. It is worth mentioning that although the ADM mass $\mathcal{E}$ does not algebraically match the mass $\mathcal{M}$ obtained using Padmanabhan's method, these expressions are numerically equivalent.

The black hole entropy can be calculated by \cite{Wald:1993nt, Hammad:2019oyb}:
\begin{eqnarray}
\notag \mathcal{S} &=& \frac{2 \pi}{\tilde{\kappa}} \int_{\Sigma} d \Sigma_{\mu \nu} Q^{\mu \nu} \\
&=& - \frac{2 \pi}{\tilde{\kappa}} \int_{\Sigma} \left( \frac{1}{2 \kappa} +\eta \phi^4 \right) \xi^a S_{a}^{\,\,\, \mu \nu} \sqrt{h} \epsilon_{\mu \nu \rho \sigma} dx^{\rho} \wedge dx^{\sigma}\,.
\end{eqnarray}
Now, we write $\epsilon_{\mu \nu \rho \sigma} = \frac{1}{2} \epsilon^{\alpha \beta} \epsilon_{\alpha \beta \mu \nu} \epsilon_{\rho \sigma}$. Also, using the identity $\epsilon^{\alpha \beta} = \epsilon^{\alpha \beta \gamma \delta} N_{\gamma} \xi_{\delta}$, where $N^{\gamma}$ is an auxiliary null vector on the horizon which satisfies $N_{\mu} \xi^{\mu} =-1$, and in addition using $\epsilon_{\alpha \beta \mu \nu} \epsilon^{\alpha \beta \gamma \delta}= - 2 (\delta_{\mu} ^{\gamma} \delta_{\nu}^{\delta}- \delta_{\nu}^{\gamma} \delta_{\mu}^{\delta})$ we arrive at
\begin{equation}
\mathcal{S}=\frac{2 \pi}{ \tilde{\kappa}} \int_{\Sigma} \left( \frac{1}{\kappa} +  2\eta \phi^4 \right) \xi^{\rho} S_{\rho} ^{\,\,\, \gamma \delta} N_{\gamma} \xi_{\delta} \sqrt{h} \epsilon_{\rho \sigma} dx^{\rho} \wedge dx^{\sigma} \,.
\end{equation}
Finally, replacing the superpotential $S_{\rho}^{\,\,\, \gamma \delta}$ given in Eq. (\ref{S}), and using $\xi^{\mu} \xi^{\nu} T_{\mu \nu \rho} =0$ and $\xi_{\mu} T^{\mu} = \tilde{\kappa}$, the black hole entropy yields 
\begin{equation} \label{entropy}
\mathcal{S} = 8 \pi^2 r_H^2 \left(\frac{1}{\kappa}+ 2\eta \psi_H^4  \right)=\frac{A}{4 G} + 4\pi f(\psi_H)A\,,
\end{equation}
this same expression for the entropy was obtained using the Padmanabhan method in (\ref{Entropy}).

\begin{figure}[ht]
\begin{center}
\includegraphics[width=0.4\textwidth]{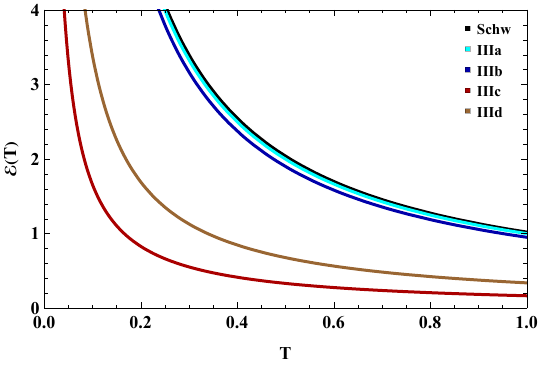}
\includegraphics[width=0.41\textwidth]{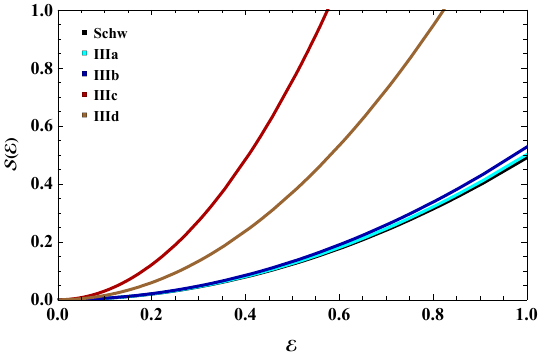}
\end{center}
\caption{The behaviour of the black hole mass as a function of the temperature $\mathcal{E}(T)$ (left panel) and the black hole entropy as a function of the mass $\mathcal{S}(\mathcal{E})$ (right panel). Here $\eta=20$.}
\label{Termn2}
\end{figure}

In the following, we consider the ADM mass given by Eq. (\ref{mass}) and the entropy in Eq. (\ref{entropy}), and we show in Fig. \ref{Termn2} the behaviour of the black hole mass as a function of temperature (left panel) and black hole entropy as a function of mass (right panel) for $\eta=20$. The solutions IIIa (zero nodes) and IIIb (one node) correspond to small deviations from the Schwarzschild case, while solutions IIIc (two nodes) and IIId (two nodes) correspond to larger deviations from the Schwarzschild case from a thermodynamic point of view. It is important to highlight that the black hole temperature decreases as the event horizon increases, and the temperature of the scalarized solutions is lower than that of the Schwarzschild solution. We observe that the mass decreases as the temperature increases, and the mass of the scalarized solutions is lower than that of the Schwarzschild solution. For a fixed value of mass, the entropy of the scalarized solutions is higher than that of the Schwarzschild black hole; therefore, the scalarized black holes are entropically favored over the Schwarzschild solution. Also, note that the solution IIIc which has the greatest value for the scalar torsion, is the solution that presents the lowest value for the black hole mass and the highest value for entropy.

Now, to study phase transitions between the scalarized and the Schwarzschild black holes, we must consider both black holes in the same canonical ensemble, i.e. at the same temperature $T$. Equaling $T$ for both black holes and by considering the free energy $\mathcal{F}$
\begin{equation}
\mathcal{F}= \mathcal{E}-T\mathcal{S} = \frac{\sqrt{a_H}}{T} \left( \frac{1 - b_1}{\kappa} -\frac{\sqrt{a_H}}{2} \left(\frac{1}{\kappa}+2 \eta \psi_H^4\right)  \right)\,.
\end{equation}
Note that for Schwarzschild black hole $\mathcal{F}=M/2$ in agreement with previous results reported \cite{Gecse:2008hj,Constantineau:2011fw}.

To study second-order phase transitions of scalarized black holes, it is known that these transitions occur at points where the heat capacity diverges. The heat capacity of the black hole is given by
\begin{equation}\label{heat}
    \mathcal{C} = \frac{\partial \mathcal{E}}{\partial T} = -\frac{ (1 - b_1)\sqrt{ a_H}}{T^2} = -\frac{\mathcal{E}}{T}\,.
\end{equation}

\begin{figure}[ht]
\begin{center}
\includegraphics[width=0.4\textwidth]{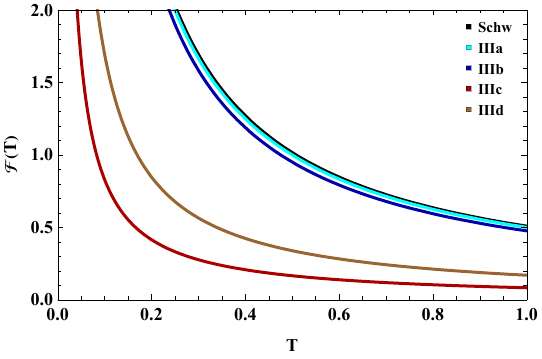}
\includegraphics[width=0.41\textwidth]{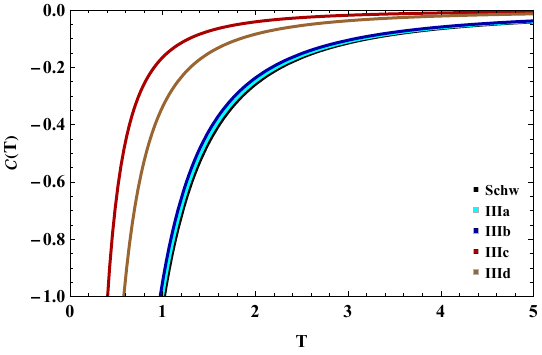}
\end{center}
\caption{The behaviour of the free energy $\mathcal{F}$ (left panel), and  the heat capacity $\mathcal{C}$ (right panel) as a function of $T$ with $\kappa=1$ and $\eta=20$.
}
\label{heatplot}
\end{figure}

In Fig. \ref{heatplot}, we plot the free energy (left panel) for the scalarized and Schwarzschild black hole, and we can observe that  there is not first order phase transitions, which is described by the characteristic swallowtail behaviour. It is worth to mention that this phase transition occurs for hyperbolic horizon in the context of hairy black holes \cite{Martinez:2010ti,Kolyvaris:2010yyf,Gonzalez:2013aca,Gonzalez:2014tga}. Interestingly enough, the free energy of the scalarized black hole solutions is lower than that of the Schwarzschild solution, so the scalarized black holes are always thermodynamically favored over Schwarzschild black holes, see the ratio of free energies $\mathcal{F}_{\text{scalarized}}/\mathcal{F}_{\text{Schwarzschild}}$ in Appendix \ref{Solutions}, Table \ref{parametersolutionsA}, for other values of the parameter $\eta$, which is always less than 1. Also, we plot the behavior of the heat capacity as a function of temperature (right panel), and it increases as the temperature rises. Note that the heat capacity is always negative; this phase is known as a meta-stable black hole. Additionally, its behavior does not exhibit divergences, meaning there are no second-order phase transitions for the solutions considered. Also, note that solution IIIc, which has the greatest value of the scalar torsion, is the solution that presents the lowest value of the free energy and the greatest value of the heat capacity.


\section{Final Remarks}
\label{conclusion}

In this work, we have considered scalar-torsion theory within the framework of Teleparallel Gravity, where the scalar field is coupled to the Weitzenb\"{o}ck invariant through a coupling function that prevents tachyonic instabilities. Using the shooting method, we have numerically demonstrated that this model leads to the formation of new black holes with scalar hair, representing a scalarization of the Schwarzschild black hole solution. We then analyzed asymptotically flat solutions for scalar fields with zero, one, and two nodes, focusing on their thermodynamics. In this analysis, we considered an energy-momentum pseudo-current, which allows for the calculation of the gravitational energy of the solutions, as well as methods developed by Padmanabhan and Wald's formalism.

 We have found two types of solutions: one described by small deviations from the Schwarzschild case, except for the scalar field, and the other by larger deviations from the Schwarzschild case. For solutions that have more than one solution with the same number of nodes, the highest value of the torsion scalar corresponds to the solution with nodes closest to the event horizon. Additionally, this solution presents the lowest value for the black hole mass and free energy, as well as the highest values for the entropy and heat capacity. Our main findings indicate that scalarized black hole solutions are metastable, with no first- or second-order phase transitions. Additionally, scalarized black holes are always preferred over the Schwarzschild black holes, as they have a lower free energy and are entropically favored.  In Appendix \ref{constraints} we explore the constraints on the coupling constant $\eta$ of the model using data from the Event Horizon Telescope (EHT) for the black hole shadows of M87* and Sgr A*. All the values of $\eta$ considered in this work meet the bounds set by the EHT data for both M87* and Sgr A*. However, the available data do not provide strong constraints on the value of $\eta$. Additionally, the angular diameter of the scalarized black hole shadow is smaller than that of the Schwarzschild black hole. 

It would be interesting to explore the dynamical stability of these solutions using the gravitational quasinormal mode spectrum. Furthermore, investigating the temporal evolution and dynamical stability of a probe scalar field would provide additional insights. Future work could also involve considering higher values of the azimuthal number \(\ell\), given the anomalous behavior of the decay rate, where the longest-lived modes are those with higher angular momentum for massless scalar fields \cite{Lagos:2020oek, Aragon:2020tvq}. Research on charged scalarized black hole solutions is also ongoing.
\\

\acknowledgments

We thank Sebastian Bahamonde for his comments and suggestions. 
This work is partially supported by ANID Chile through FONDECYT Grant Nº 1220871  (P.A.G., J.R., and Y. V.). P.A.G. would like to thank the Facultad de Ciencias, Universidad de La Serena for its hospitality.


\appendix
\section{Parameters Values for Scalarized Black Hole Solutions}
\label{Solutions}

In the following table, we show the values of the parameters $\eta$, $n$, $a_H$, $\psi_H$, $\psi_\infty$, $b_1$, and $\mathcal{F}_{\text{scalarized}}/\mathcal{F}_{\text{Schwarzschild}}$ for both the Schwarzschild and scalarized black hole solutions. Here, $\mathcal{F}_{\text{scalarized}}$ corresponds to the free energy for the scalarized solution and $\mathcal{F}_{\text{Schwarzschild}}$ corresponds to the free energy for the Schwarzschild solution. Note that the ratio $\mathcal{F}_{\text{scalarized}}/\mathcal{F}_{\text{Schwarzschild}}$ is always less than 1, indicating that scalarized black holes are always thermodynamically favored over Schwarzschild black holes.

\begin{table}[ht]
  \centering
   \scalebox{0.8} {
  \begin{tabular}{|c|c|c|c|c|c|c|c|}
    \hline
    \textbf{Type of BH} & $\boldsymbol{\eta}$ & \textbf{Number of nodes} & $\boldsymbol{a_H}$ & $\boldsymbol{\psi_H}$ & $\boldsymbol{\psi_\infty}$ & $\boldsymbol{b_1}$ & $\mathcal{F}_{\textbf{scalarized}}/\mathcal{F}_{\textbf{Schwarzschild}}$ \\
    \hline
    Schwarzschild & 0 & {} & 1 & 0 & 0 & 0 & {} \\
    \hline
      \multirow{2}{*}{Scalarized} & \multirow{2}{*}{0.060}  & 0 & $4.416 \times 10^{-7}$ & 11.154 & 0.450 & -0.235 & $8.207 \times 10^{-4}$ \\
    \cline{3-8}
     &  & 0 & 0.493 & 1.536 & 0.687 & -0.171 & 0.822 \\ 
    \hline
      \multirow{2}{*}{Scalarized} & \multirow{2}{*}{0.065}  & 0 & $1.153 \times 10^{-6}$ & 9.672 & 0.418 & -0.222 & 0.001 \\
    \cline{3-8}
    &  & 0 & 0.521 & 1.468 & 0.663 & -0.159 & 0.836 \\
    \hline
      Scalarized & 1  & 0 & 0.959 & 0.354 & 0.176 & -0.010 & 0.998\\
    \hline
    
    \multirow{3}{*}{Scalarized} &   \multirow{3}{*}{2} & 0 & 0.979 & 0.250 & 0.125 & -0.005 & 0.994 \\
    \cline{3-8}
    &  & 1 & 0.004 & 1.410 & -0.540 & -0.070 & 0.068 \\
    \cline{3-8}
    &  & 1 & 0.205 & 0.764 & -0.719 & -0.068 & 0.482 \\
    \hline
    \multirow{3}{*}{Scalarized} &   \multirow{3}{*}{5} & 0 & 0.991 & 0.158 & 0.079 & -0.002 & 0.997\\
    \cline{3-8}
    &  & 1 & $1.083\times 10^{-4}$ & 1.771 & -0.334 & -0.035 & 0.010 \\
    \cline{3-8}
    &  & 1 & 0.606 & 0.422 & -0.509 & -0.028 & 0.802 \\
    \hline

     \multirow{2}{*}{Scalarized} &   \multirow{2}{*}{10} & 0 & 0.995 & 0.111 & 0.055 & -0.001 & 0.998 \\
    \cline{3-8}
    &  & 1 & 0.788 & 0.290 & -0.370 & -0.014 & 0.900 \\
    \hline

     \multirow{2}{*}{Scalarized} &   \multirow{2}{*}{15} & 0 & 0.997 & 0.091 & 0.045 & $-6.945\times 10^{-4}$ & 0.999 \\
    \cline{3-8}
    &  & 1 & 0.855 & 0.234 & -0.305 & -0.009 & 0.934 \\
    \hline

    \multirow{4}{*}{Scalarized} &  \multirow{4}{*}{20} & 0 & 0.998 & 0.079 & 0.040 & -0.0005 & 0.999 \\
    \cline{3-8}
    &  & 1 & 0.890 & 0.203 & -0.265 & -0.007 & 0.949 \\
    \cline{3-8}
    &  & 2 & 0.026 & 0.607 & 0.649 & -0.032 & 0.165 \\
    \cline{3-8}
    &  & 2 & 0.108 & 0.481 & 0.704 & -0.032 & 0.339 \\
    \hline

     \multirow{4}{*}{Scalarized} &  \multirow{4}{*}{25} & 0 & 0.998 & 0.070 & 0.035 & $-4.167\times 10^{-4}$ & 0.999 \\
    \cline{3-8}
    &  & 1 & 0.911 & 0.180 & -0.237 & -0.005 & 0.959 \\
    \cline{3-8}
    &  & 2 & 0.006 & 0.700 & 0.549 & -0.026 & 0.081 \\
    \cline{3-8}
    &  & 2 & 0.255 & 0.378 & 0.670 & -0.026 & 0.521 \\
    \hline

  \end{tabular}}
  \caption{Values of the parameters $\eta$, $n$, $a_H$, $\psi_H$, $\psi_\infty$, and $b_1$ for the Schwarzschild and scalarized black hole solutions, and the ratio of free energies $\mathcal{F}_{\text{scalarized}}/\mathcal{F}_{\text{Schwarzschild}}$.}
  \label{parametersolutionsA}
\end{table}

\section{Using EHT data of M87* and Sgr A* to constrain $\eta$}
\label{constraints}

In this appendix, we explore the constraints on the coupling constant $\eta$ of the model using data from the Event Horizon Telescope (EHT) for the black hole shadows of M87* and Sgr A*. A review of analytical studies of black hole shadows was performed in 
\cite{Perlick:2021aok}. The EHT data have been used as a probe to distinguish black holes with and without spontaneous scalarization in Einstein-scalar-Gauss-Bonnet theories in \cite{Wang:2024lte}.

In teleparallel gravity, there are no geodesics, only force equations analogous to the Lorentz force equation of electrodynamics; however, the teleparallel force equation coincides with the geodesic equation of General Relativity \cite{JGPereira}, and for the spherically symmetric metric \eqref{metric1}, the equation of motion for a photon moving in the equatorial plane is given by
\begin{equation}
\frac{dr}{d \varphi} = \pm r^2 \sqrt{\frac{B(r)}{A(r)} \left( \frac{1}{b^2}- V_{\text{eff}} (r) \right)}\,,
\end{equation}
where $b = L/E$ is the impact parameter, $E$ and $L$ are the conserved energy and angular momentum of the photon, respectively. The $\pm$ sign refers to the photons moving counterclockwise (positive) or clockwise (negative) on the equatorial plane. The effective potential is given by
\begin{equation}
V_{\text{eff}}(r) = \frac{A(r)}{r^2}\,,
\end{equation}
and the radius of the photon sphere, $r_{PS}$, is determined by the conditions $dr/d\varphi=0$ and $d^2r/d\varphi^2 =0$, which yields
\begin{equation} \label{equa}
V_{\text{eff}} (r_{PS}) = \frac{1}{b_{PS}^2} \,,  \quad \quad  V'_{\text{eff}} (r_{PS}) = 0\,,
\end{equation}
where $b_{PS}$ is the critical impact parameter, an important quantity related to the black hole shadow, as it separates the captured orbits from the flyby orbits of the incident light rays. On the other hand, the angular diameter $\theta_{Sh}$ of the black hole shadow (angular size of the shadow), as measured by a distant observer, is given approximated by
\begin{equation} \label{angulo}
\theta_{Sh} \approx 2 \frac{b_{PS}}{r_0}\,,
\end{equation}
where $r_0$ is the distance between the black hole and the observer. To constrain the parameter $\eta$, we use the EHT data for the black hole shadows of M87* and Sgr A*. For M87*, with a mass of $M=6.5 \times 10^9 M_{\odot}$ and a distance of $r_0=16.8$ Mpc, the angular diameter is constrained to lie between $29.32$ $\mu as$ and $51.06$ $\mu as$ \cite{EventHorizonTelescope:2019dse, EventHorizonTelescope:2021dqv}. For Sgr A*, with a mass of $M=4.0 \times 10^6 M_{\odot}$ and a distance of $r_0 = 8.15$ kpc, the angular diameter is measured to be $\theta_{Sh} = 48.7 \pm 7$ $\mu as$ \cite{EventHorizonTelescope:2022wkp, EventHorizonTelescope:2022xqj}.

Next, we express the effective potential in terms of the coordinate $z$ as $V_{\text{eff}}= \frac{1}{r_H^2} A(z) (1-z)^2$. 
By determining the value of $z$ where the potential reaches its maximum, and using the ADM mass of the black hole $\mathcal{E}= r_H (1-b_1) c^2/2G$ to compute $r_H$, we can obtain $b_{PS}$ from Eq. (\ref{equa}). The angular diameter is then computed from Eq. (\ref{angulo}).  The angular diameters for the scalarized black hole solutions, listed in Table \ref{parametersolutionsA} of Appendix \ref{Solutions}, are plotted in Figure (\ref{datos}). From these, we observe that all the values of the coupling constant $\eta$ considered in this work satisfy the bounds set by the EHT data for both M87* and Sgr A*.

\begin{figure}[h!]
	
    \centering
         \includegraphics[width=0.45\linewidth]{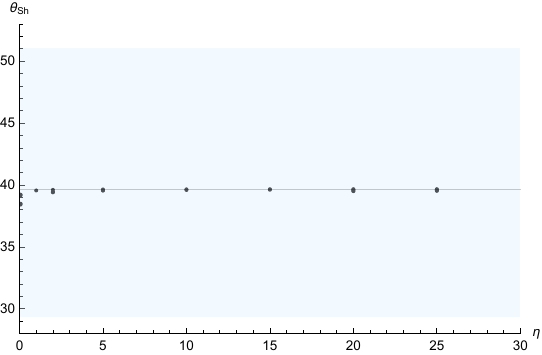}
         \includegraphics[width=0.45\linewidth]{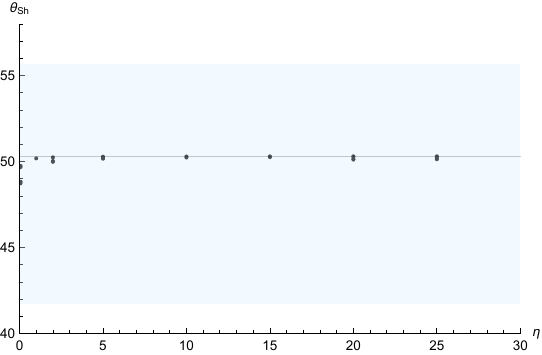}
         \caption{Angular diameter $\theta_{Sh}$ vs. $\eta$. The dots represent the angular diameters for the scalarized black hole solutions listed in the Table \ref{parametersolutionsA} of Appendix \ref{Solutions}, while the horizontal line corresponds to the value for the Schwarschild black hole. The shaded region denotes the observational bounds. The left panel shows the results for M87*, while the right panel shows the corresponding results for Sgr A*.}
           \label{datos}
\end{figure}



\begin{thebibliography}{99}

\bibitem{chaostro}
M. Ostrogradski,
Memoires sur les equations differentielles relatives au probleme des
isoperimetres,
 Mem. Ac. St. Petersbourg 4, 385 (1850).



\bibitem{Will:2005va}
C.~M.~Will,
``The Confrontation between general relativity and experiment,''
Living Rev. Rel. \textbf{9}, 3 (2006)
[arXiv:gr-qc/0510072 [gr-qc]].

 
\bibitem{Unzicker:2005in}
A.~Unzicker and T.~Case,
``Translation of Einstein's attempt of a unified field theory with teleparallelism,''
[arXiv:physics/0503046 [physics]].

\bibitem{TEGR}
   C. M\"{o}ller,
   Conservation laws and absolute parallelism in general relativity,
    Mat. Fys. Skr. Dan. Vid. Selsk. {\bf 1}, 3
(1961).

\bibitem{TEGR22}
  C. Pellegrini and J. Plebanski,
   Tetrad fields and gravitational fields,
   Mat. Fys. Skr. Dan. Vid. Selsk. {\bf 2}, 1
(1963).

\bibitem{Hayashi:1979qx}
  K.~Hayashi and T.~Shirafuji,
     New general relativity,
  Phys.\ Rev.\  D {\bf 19}, 3524 (1979)
  [Addendum-ibid.\  D {\bf 24}, 3312 (1982)].



    \bibitem{JGPereira}
R. Aldrovandi and J. G. Pereira,  Teleparallel Gravity: An Introduction,
Springer, Dordrecht (2013).


\bibitem{Arcos:2004tzt}
H.~I.~Arcos and J.~G.~Pereira,
``Torsion gravity: A Reappraisal,''
Int. J. Mod. Phys. D \textbf{13}, 2193-2240 (2004)
[arXiv:gr-qc/0501017 [gr-qc]].




\bibitem{Maluf:2013gaa}
J.~W.~Maluf,
``The teleparallel equivalent of general relativity,''
Annalen Phys. \textbf{525}, 339-357 (2013)
[arXiv:1303.3897 [gr-qc]].


\bibitem{Pereira:2013qza}
J.~G.~Pereira,
``Teleparallelism: A New Insight Into Gravity,''
Chapter in ``Springer Handbook of Spacetime'', ed. by A. Ashtekar and V. Petrov (Springer, Dordrecht, 2014)
[arXiv:1302.6983 [gr-qc]].

\bibitem{Capozziello:2022zzh}
S.~Capozziello, V.~De Falco and C.~Ferrara,
``Comparing equivalent gravities: common features and differences,''
Eur. Phys. J. C \textbf{82}, no.10, 865 (2022)
[arXiv:2208.03011 [gr-qc]].

\bibitem{Wang:2011xf}
T.~Wang,
``Static Solutions with Spherical Symmetry in f(T) Theories,''
Phys. Rev. D \textbf{84}, 024042 (2011)
[arXiv:1102.4410 [gr-qc]].

\bibitem{Miao:2011ki}
R.~X.~Miao, M.~Li and Y.~G.~Miao,
``Violation of the first law of black hole thermodynamics in $f(T)$ gravity,''
JCAP \textbf{11}, 033 (2011)
[arXiv:1107.0515 [hep-th]].

\bibitem{Gonzalez:2011dr}
P.~A.~Gonzalez, E.~N.~Saridakis and Y.~Vasquez,
``Circularly symmetric solutions in three-dimensional Teleparallel, f(T) and Maxwell-f(T) gravity,''
JHEP \textbf{07}, 053 (2012)
[arXiv:1110.4024 [gr-qc]].

\bibitem{Capozziello:2012zj}
S.~Capozziello, P.~A.~Gonzalez, E.~N.~Saridakis and Y.~Vasquez,
``Exact charged black-hole solutions in D-dimensional f(T) gravity: torsion vs curvature analysis,''
JHEP \textbf{02}, 039 (2013)
[arXiv:1210.1098 [hep-th]].

\bibitem{Atazadeh:2012am}
K.~Atazadeh and M.~Mousavi,
``Vacuum spherically symmetric solutions in $f(T)$ gravity,''
Eur. Phys. J. C \textbf{73}, no.1, 2272 (2013)
[arXiv:1212.3764 [gr-qc]].

\bibitem{Calza:2023hhi}
M.~Calz\'a and L.~Sebastiani,
``A class of static spherically symmetric solutions in f(T)-gravity,''
Eur. Phys. J. C \textbf{84} (2024) no.5, 476
[arXiv:2309.04536 [gr-qc]].




\bibitem{Ferraro:2006jd}
R.~Ferraro and F.~Fiorini,
``Modified teleparallel gravity: Inflation without inflaton,''
Phys. Rev. D \textbf{75}, 084031 (2007)
[arXiv:gr-qc/0610067 [gr-qc]].

\bibitem{Linder:2010py}
E.~V.~Linder,
``Einstein's Other Gravity and the Acceleration of the Universe,''
Phys. Rev. D \textbf{81}, 127301 (2010)
[erratum: Phys. Rev. D \textbf{82}, 109902 (2010)]
[arXiv:1005.3039 [astro-ph.CO]].

\bibitem{Li:2013xea}
J.~T.~Li, C.~C.~Lee and C.~Q.~Geng,
``Einstein Static Universe in Exponential $f(T)$ Gravity,''
Eur. Phys. J. C \textbf{73}, no.2, 2315 (2013)
[arXiv:1302.2688 [gr-qc]].

\bibitem{Kofinas:2014aka}
G.~Kofinas, G.~Leon and E.~N.~Saridakis,
``Dynamical behavior in $f(T,T_G)$ cosmology,''
Class. Quant. Grav. \textbf{31}, 175011 (2014)
[arXiv:1404.7100 [gr-qc]].

\bibitem{Kofinas:2014daa}
G.~Kofinas and E.~N.~Saridakis,
``Cosmological applications of $F(T,T_G)$ gravity,''
Phys. Rev. D \textbf{90}, 084045 (2014)
[arXiv:1408.0107 [gr-qc]].

\bibitem{Fujii}
 Y. Fujii, K. Maeda,
 "The scalar-tensor theory of gravitation"
 (Cambridge University Press, 2007).

\bibitem{BBMB}
  N. Bocharova, K. Bronnikov and V. Melnikov, Vestn. Mosk.
  Univ. Fiz. Astron. \textbf{6}, 706 (1970);\\
  J.~D.~Bekenstein,
  ``Exact solutions of Einstein conformal scalar equations,''
  Annals Phys.\  {\bf 82}, 535 (1974);\\
  J.~D.~Bekenstein, ``Black Holes With Scalar Charge,''
  Annals Phys.\ \textbf{91}, 75 (1975). 
  
\bibitem{bronnikov}
  K.~A.~Bronnikov and Y.~N.~Kireev,
  ``Instability of Black Holes with Scalar Charge,''
  Phys.\ Lett.\ A {\bf 67}, 95 (1978).
  
\bibitem{Martinez:1996gn}
C.~Martinez and J.~Zanelli,
``Conformally dressed black hole in (2+1)-dimensions,''
Phys. Rev. D \textbf{54}, 3830-3833 (1996)
[arXiv:gr-qc/9604021 [gr-qc]].

\bibitem{Banados:1992wn}
M.~Banados, C.~Teitelboim and J.~Zanelli,
``The Black hole in three-dimensional space-time,''
Phys. Rev. Lett. \textbf{69}, 1849-1851 (1992)
[arXiv:hep-th/9204099 [hep-th]].

\bibitem{Martinez:2002ru}
C.~Martinez, R.~Troncoso and J.~Zanelli,
``De Sitter black hole with a conformally coupled scalar field in four-dimensions,''
Phys. Rev. D \textbf{67}, 024008 (2003)
[arXiv:hep-th/0205319 [hep-th]].

\bibitem{Winstanley:2002jt}
E.~Winstanley,
``On the existence of conformally coupled scalar field hair for black holes in (anti-)de Sitter space,''
Found. Phys. \textbf{33}, 111-143 (2003)
[arXiv:gr-qc/0205092 [gr-qc]].

\bibitem{Kolyvaris:2010yyf}
T.~Kolyvaris, G.~Koutsoumbas, E.~Papantonopoulos and G.~Siopsis,
``A New Class of Exact Hairy Black Hole Solutions,''
Gen. Rel. Grav. \textbf{43}, 163-180 (2011)
[arXiv:0911.1711 [hep-th]].

\bibitem{Charmousis:2014zaa}
C.~Charmousis, T.~Kolyvaris, E.~Papantonopoulos and M.~Tsoukalas,
``Black Holes in Bi-scalar Extensions of Horndeski Theories,''
JHEP \textbf{07}, 085 (2014)
[arXiv:1404.1024 [gr-qc]].

\bibitem{Doneva_2018a}
  D.~D.~Doneva and S.~S.~Yazadjiev,
 ``New Gauss-Bonnet Black Holes with Curvature-Induced Scalarization in Extended Scalar-Tensor Theories,''
  Phys.\ Rev.\ Lett.\  {\bf 120}, no. 13, 131103 (2018)
  [arXiv:1711.01187 [gr-qc]].

\bibitem{Silva_2018}
  H.~O.~Silva, J.~Sakstein, L.~Gualtieri, T.~P.~Sotiriou and E.~Berti,
 ``Spontaneous scalarization of black holes and compact stars from a Gauss-Bonnet coupling,''
  Phys.\ Rev.\ Lett.\  {\bf 120}, no. 13, 131104 (2018)
  [arXiv:1711.02080 [gr-qc]].

 

\bibitem{Chen:2006ge}
  C.~M.~Chen, D.~V.~Gal'tsov and D.~G.~Orlov,
``Extremal black holes in D=4 Gauss-Bonnet gravity,''
  Phys.\ Rev.\ D {\bf 75}, 084030 (2007)
  [hep-th/0701004].


\bibitem{Antoniou_2018}
  G.~Antoniou, A.~Bakopoulos and P.~Kanti,
  ``Evasion of No-Hair Theorems and Novel Black-Hole Solutions in Gauss-Bonnet Theories,''
  Phys.\ Rev.\ Lett.\  {\bf 120}, no. 13, 131102 (2018)
  [arXiv:1711.03390 [hep-th]].

\bibitem{Antoniou_2018a}
  G.~Antoniou, A.~Bakopoulos and P.~Kanti,
 ``Black-Hole Solutions with Scalar Hair in Einstein-Scalar-Gauss-Bonnet Theories,''
  Phys.\ Rev.\ D {\bf 97}, no. 8, 084037 (2018)
  [arXiv:1711.07431 [hep-th]].
  
\bibitem{Cunha:2019dwb} 
  P.~V.~P.~Cunha, C.~A.~R.~Herdeiro and E.~Radu,
  ``Spontaneously Scalarized Kerr Black Holes in Extended Scalar-Tensor–Gauss-Bonnet Gravity,''
  Phys.\ Rev.\ Lett.\  {\bf 123}, no. 1, 011101 (2019)
  [arXiv:1904.09997 [gr-qc]].
  
  
  
  
\bibitem{Macedo:2019sem} 
  C.~F.~B.~Macedo, J.~Sakstein, E.~Berti, L.~Gualtieri, H.~O.~Silva and T.~P.~Sotiriou,
  ``Self-interactions and Spontaneous Black Hole Scalarization,''
  Phys.\ Rev.\ D {\bf 99}, no. 10, 104041 (2019)
  [arXiv:1903.06784 [gr-qc]].
  
  
  
\bibitem{Peng:2020znl}
  Y.~Peng,
  ``Spontaneous scalarization of Gauss-Bonnet black holes surrounded by massive scalar fields,''
  Phys.\ Lett.\ B {\bf 807} (2020) 135569
  [arXiv:2004.12566 [gr-qc]].
  
\bibitem{Bekenstein:1992pj}
J.~D.~Bekenstein,
``The Relation between physical and gravitational geometry,''
Phys. Rev. D \textbf{48} (1993), 3641-3647
[arXiv:gr-qc/9211017 [gr-qc]].
  
  
\bibitem{Bettoni:2013diz} 
  D.~Bettoni and S.~Liberati,
  ``Disformal invariance of second order scalar-tensor theories: Framing the Horndeski action,''
  Phys.\ Rev.\ D {\bf 88}, 084020 (2013)
  [arXiv:1306.6724 [gr-qc]].
  
  
  
\bibitem{Minamitsuji:2016hkk} 
  M.~Minamitsuji and H.~O.~Silva,
  ``Relativistic stars in scalar-tensor theories with disformal coupling,''
  Phys.\ Rev.\ D {\bf 93}, no. 12, 124041 (2016)
  [arXiv:1604.07742 [gr-qc]].
  
 
    
  
\bibitem{Damour:1993hw} 
  T.~Damour and G.~Esposito-Farese,
  ``Nonperturbative strong field effects in tensor - scalar theories of gravitation,''
  Phys.\ Rev.\ Lett.\  {\bf 70}, 2220 (1993).
  
\bibitem{Dcited}
P. G. Bergmann, Int. J. Theor. Phys. {\bf{1}}, 25 (1968); K. Nordtvedt, Astrophys. J. {\bf{161}}, 1059 (1970); R. V. Wagoner, Phys. Rev. D. {\bf{1}}, 3209 (1970).

\bibitem{Dcited2}
T. Damour and J.H. Taylor, Phys. Rev. D 45, 1840 (1992).
   
\bibitem{Chen:2015zmx} 
  P.~Chen, T.~Suyama and J.~Yokoyama,
  ``Spontaneous scalarization: asymmetron as dark matter,''
  Phys.\ Rev.\ D {\bf 92}, 124016 (2015)
  [arXiv:1508.01384 [gr-qc]].
  
  
\bibitem{Ramazanoglu:2016kul}
F.~M.~Ramazanoğlu and F.~Pretorius,
``Spontaneous Scalarization with Massive Fields,''
Phys. Rev. D \textbf{93} (2016) no.6, 064005
[arXiv:1601.07475 [gr-qc]].


  
\bibitem{Berti:2015itd} 
  E.~Berti {\it et al.},
  ``Testing General Relativity with Present and Future Astrophysical Observations,''
  Class.\ Quant.\ Grav.\  {\bf 32}, 243001 (2015)
  [arXiv:1501.07274 [gr-qc]].
  
\bibitem{Brihaye:2018bgc} 
  Y.~Brihaye, C.~Herdeiro and E.~Radu,
  ``The scalarised Schwarzschild-NUT spacetime,''
  Phys.\ Lett.\ B {\bf 788}, 295 (2019)
  [arXiv:1810.09560 [gr-qc]].
  
  
  
\bibitem{Herdeiro:2018wub} 
  C.~A.~R.~Herdeiro, E.~Radu, N.~Sanchis-Gual and J.~A.~Font,
  ``Spontaneous Scalarization of Charged Black Holes,''
  Phys.\ Rev.\ Lett.\  {\bf 121}, no. 10, 101102 (2018)
  [arXiv:1806.05190 [gr-qc]].
  
\bibitem{Kiorpelidi:2023jjw}
S.~Kiorpelidi, T.~Karakasis, G.~Koutsoumbas and E.~Papantonopoulos,
``Scalarization of the Reissner-Nordstr\"om black hole with higher derivative gauge field corrections,''
Phys. Rev. D \textbf{109}, no.2, 024033 (2024)
[arXiv:2311.10858 [gr-qc]].
  
  
\bibitem{Minamitsuji:2019iwp} 
  M.~Minamitsuji and T.~Ikeda,
  ``Spontaneous scalarization of black holes in the Horndeski theory,''
  Phys.\ Rev.\ D {\bf 99}, no. 10, 104069 (2019)
  [arXiv:1904.06572 [gr-qc]].

  
\bibitem{Dima:2020yac}
A.~Dima, E.~Barausse, N.~Franchini and T.~P.~Sotiriou,
``Spin-induced black hole spontaneous scalarization,''
Phys. Rev. Lett. \textbf{125} (2020) no.23, 231101
[arXiv:2006.03095 [gr-qc]].

\bibitem{Guo:2020sdu}
  H.~Guo, S.~Kiorpelidi, X.~M.~Kuang, E.~Papantonopoulos, B.~Wang and J.~P.~Wu,
  ``Spontaneous holographic scalarization of black holes in Einstein-scalar-Gauss-Bonnet theories,''
  Phys.\ Rev.\ D {\bf 102} (2020) no.8,  084029
  [arXiv:2006.10659 [hep-th]].
  
\bibitem{Doneva:2021dcc}
  D.~D.~Doneva and S.~S.~Yazadjiev,
  ``Spontaneously scalarized black holes in dynamical Chern-Simons gravity: dynamics and equilibrium solutions,''
  Phys.\ Rev.\ D {\bf 103} (2021) no.8,  083007
  [arXiv:2102.03940 [gr-qc]].
  
\bibitem{Zhang:2022sgt}
S.~J.~Zhang, B.~Wang, E.~Papantonopoulos and A.~Wang,
``Magnetic-induced spontaneous scalarization in dynamical Chern\textendash{}Simons gravity,''
Eur. Phys. J. C \textbf{83}, no.1, 97 (2023)
[arXiv:2209.02268 [gr-qc]].

\bibitem{Chatzifotis:2022mob}
N.~Chatzifotis, P.~Dorlis, N.~E.~Mavromatos and E.~Papantonopoulos,
``Scalarization of Chern-Simons-Kerr black hole solutions and wormholes,''
Phys. Rev. D \textbf{105}, no.8, 084051 (2022)
[arXiv:2202.03496 [gr-qc]].



\bibitem{Ripley:2020vpk}
J.~L.~Ripley and F.~Pretorius,
``Dynamics of a $\mathbb Z_2$ symmetric EdGB gravity in spherical symmetry,''
Class. Quant. Grav. \textbf{37} (2020) no.15, 155003
[arXiv:2005.05417 [gr-qc]].


\bibitem{Silva:2020omi}
H.~O.~Silva, H.~Witek, M.~Elley and N.~Yunes,
``Dynamical Descalarization in Binary Black Hole Mergers,''
Phys. Rev. Lett. \textbf{127} (2021) no.3, 031101
[arXiv:2012.10436 [gr-qc]].



\bibitem{Doneva:2021dqn}
D.~D.~Doneva and S.~S.~Yazadjiev,
``Dynamics of the nonrotating and rotating black hole scalarization,''
Phys. Rev. D \textbf{103} (2021) no.6, 064024
[arXiv:2101.03514 [gr-qc]].


\bibitem{Kuan:2021lol}
H.~J.~Kuan, D.~D.~Doneva and S.~S.~Yazadjiev,
``Dynamical Formation of Scalarized Black Holes and Neutron Stars through Stellar Core Collapse,''
Phys. Rev. Lett. \textbf{127} (2021) no.16, 161103
[arXiv:2103.11999 [gr-qc]].


\bibitem{East:2021bqk}
W.~E.~East and J.~L.~Ripley,
``Dynamics of Spontaneous Black Hole Scalarization and Mergers in Einstein-Scalar-Gauss-Bonnet Gravity,''
Phys. Rev. Lett. \textbf{127} (2021) no.10, 101102
[arXiv:2105.08571 [gr-qc]].


\bibitem{Doneva:2021tvn}
D.~D.~Doneva and S.~S.~Yazadjiev,
``Beyond the spontaneous scalarization: New fully nonlinear mechanism for the formation of scalarized black holes and its dynamical development,''
Phys. Rev. D \textbf{105} (2022) no.4, L041502
[arXiv:2107.01738 [gr-qc]].


\bibitem{Salgado:2005pg} 
  P.~Salgado, G.~Rubilar, J.~Crisostomo and S.~del Campo,
  ``A note about teleparallel supergravity,''
  Eur.\ Phys.\ J.\ C {\bf 44}, 587 (2005).
  
\bibitem{Geng:2011aj}
  C.~Q.~Geng, C.~C.~Lee, E.~N.~Saridakis and Y.~P.~Wu,
  ``'Teleparallel' Dark Energy,''
  Phys.\ Lett.\ B {\bf 704} (2011) 384
  [arXiv:1109.1092 [hep-th]].
  

  
\bibitem{Gonzalez:2014pwa}
P.~A.~Gonz\'alez, J.~Saavedra and Y.~V\'asquez,
``Three-Dimensional Hairy Black Holes in Teleparallel Gravity,''
Astrophys. Space Sci. \textbf{357} (2015) no.2, 143
[arXiv:1411.2193 [gr-qc]].
   
\bibitem{Kofinas:2015hla}
G.~Kofinas, E.~Papantonopoulos and E.~N.~Saridakis,
``Self-Gravitating Spherically Symmetric Solutions in Scalar-Torsion Theories,''
Phys. Rev. D \textbf{91} (2015) no.10, 104034
[arXiv:1501.00365 [gr-qc]].

\bibitem{Kofinas:2015zaa} 
  G.~Kofinas,
  ``Hyperscaling violating black holes in scalar-torsion theories,''
  Phys.\ Rev.\ D {\bf 92}, no. 8, 084022 (2015)
  [arXiv:1507.07434 [hep-th]].
  



  \bibitem{Hohmann:2018vle} 
  M.~Hohmann,
  ``Scalar-torsion theories of gravity I: general formalism and conformal transformations,''
  Phys.\ Rev.\ D {\bf 98}, no. 6, 064002 (2018)
  [arXiv:1801.06528 [gr-qc]].

 
  
  \bibitem{Hohmann:2018dqh} 
  M.~Hohmann and C.~Pfeifer,
  ``Scalar-torsion theories of gravity II: $L(T, X, Y, \phi)$ theory,''
  Phys.\ Rev.\ D {\bf 98}, no. 6, 064003 (2018)
  [arXiv:1801.06536 [gr-qc]].
  
  \bibitem{Hohmann:2018ijr} 
  M.~Hohmann,
  ``Scalar-torsion theories of gravity III: analogue of scalar-tensor gravity and conformal invariants,''
  Phys.\ Rev.\ D {\bf 98}, no. 6, 064004 (2018)
  [arXiv:1801.06531 [gr-qc]].
 
 

\bibitem{Geng:2014nfa}
C.~Q.~Geng, C.~Lai, L.~W.~Luo and H.~H.~Tseng,
``Kaluza\textendash{}Klein theory for teleparallel gravity,''
Phys. Lett. B \textbf{737} (2014), 248-250
[arXiv:1409.1018 [gr-qc]].


  
  
  
  
\bibitem{Kofinas:2014owa}
  G.~Kofinas and E.~N.~Saridakis,
  ``Teleparallel equivalent of Gauss-Bonnet gravity and its modifications,''
  Phys.\ Rev.\ D {\bf 90} (2014) 8,  084044
  [arXiv:1404.2249 [gr-qc]].
  


 
\bibitem{Gonzalez:2015sha}
P.~A.~Gonzalez and Y.~Vasquez,
``Teleparallel Equivalent of Lovelock Gravity,''
Phys. Rev. D \textbf{92} (2015) no.12, 124023
[arXiv:1508.01174 [hep-th]].
  
\bibitem{Gonzalez:2019tky}
P.~A.~Gonz\'alez, S.~Reyes and Y.~V\'asquez,
``Teleparallel Equivalent of Lovelock Gravity, Generalizations and Cosmological Applications,''
JCAP \textbf{07} (2019), 040
[arXiv:1905.07633 [gr-qc]].


\bibitem{Astudillo-Neira:2017anx} 
  N.~Astudillo-Neira and P.~Salgado,
  ``Teleparallel equivalent of higher dimensional gravity theories,''
  [arXiv:1703.07831 [hep-th]].
  
  
\bibitem{Bahamonde:2017wwk}
S.~Bahamonde, C.~G.~Böhmer and M.~Krššák,
``New classes of modified teleparallel gravity models,''
Phys. Lett. B \textbf{775} (2017), 37-43
[arXiv:1706.04920 [gr-qc]].

\bibitem{Bahamonde:2022chq}
S.~Bahamonde, D.~D.~Doneva, L.~Ducobu, C.~Pfeifer and S.~S.~Yazadjiev,
``Spontaneous scalarization of black holes in Gauss-Bonnet teleparallel gravity,''
Phys. Rev. D \textbf{107} (2023) no.10, 104013
[arXiv:2212.07653 [gr-qc]].


\bibitem{Hehl:1994ue}
F.~W.~Hehl, J.~D.~McCrea, E.~W.~Mielke and Y.~Ne'eman,
``Metric affine gauge theory of gravity: Field equations, Noether identities, world spinors, and breaking of dilation invariance,''
Phys. Rept. \textbf{258}, 1-171 (1995)
[arXiv:gr-qc/9402012 [gr-qc]].



\bibitem{Geng:2011ka}
C.~Q.~Geng, C.~C.~Lee and E.~N.~Saridakis,
``Observational Constraints on Teleparallel Dark Energy,''
JCAP \textbf{01} (2012), 002
[arXiv:1110.0913 [astro-ph.CO]].


\bibitem{Staykov:2022uwq}
K.~V.~Staykov and D.~D.~Doneva,
``Multiscalar Gauss-Bonnet gravity: Scalarized black holes beyond spontaneous scalarization,''
Phys. Rev. D \textbf{106}, no.10, 104064 (2022)
[arXiv:2209.01038 [gr-qc]].


\bibitem{Bahamonde:2021srr}
S.~Bahamonde, A.~Golovnev, M.~J.~Guzm\'an, J.~L.~Said and C.~Pfeifer,
``Black holes in f(T,B) gravity: exact and perturbed solutions,''
JCAP \textbf{01} (2022) no.01, 037
[arXiv:2110.04087 [gr-qc]].

\bibitem{Bahamonde:2022lvh}
S.~Bahamonde, L.~Ducobu and C.~Pfeifer,
``Scalarized black holes in teleparallel gravity,''
JCAP \textbf{04} (2022) no.04, 018
[arXiv:2201.11445 [gr-qc]].








\bibitem{stoer1980introduction}
Stoer, Josef and Bulirsch, Roland and Bartels, R and Gautschi, Walter and Witzgall, Christoph, ``Introduction to numerical analysis'', Springer (1980).

\bibitem{press2007numerical}
Press, William H, ``Numerical recipes 3rd edition: The art of scientific computing'', Cambridge university press (2007).

\bibitem{Padmanabhan:2012gx}
T.~Padmanabhan,
``Emergent perspective of Gravity and Dark Energy,''
Res. Astron. Astrophys. \textbf{12} (2012), 891-916
[arXiv:1207.0505 [astro-ph.CO]].


\bibitem{Fiorini:2023axr}
F.~Fiorini, P.~A.~Gonz\'alez and Y.~V\'asquez,
``Reference frames and black hole thermodynamics,''
JCAP \textbf{12} (2023), 033
[arXiv:2309.06293 [gr-qc]].

\bibitem{Bras}
V.~C.~de Andrade, L.~C.~T.~Guillen and J.~G.~Pereira,
``Gravitational energy momentum density in teleparallel gravity,''
Phys. Rev. Lett. \textbf{84} (2000) 4533.
[arXiv:gr-qc/0003100 [gr-qc]].



\bibitem{Maluf1}
J.~W.~Maluf,
``The Gravitational energy-momentum tensor and the gravitational pressure,''
Annalen Phys. \textbf{14} (2005) 723.
[arXiv:gr-qc/0504077 [gr-qc]].


\bibitem{Maluf2}
J.~W.~Maluf, F.~F.~Faria and S.~C.~Ulhoa,
``On reference frames in spacetime and gravitational energy in freely falling frames,''
Class. Quant. Grav. \textbf{24} (2007) 2743.
[arXiv:0704.0986 [gr-qc]].

\bibitem{Wald:1993nt}
R.~M.~Wald,
``Black hole entropy is the Noether charge,''
Phys. Rev. D \textbf{48}, no.8, R3427-R3431 (1993)
[arXiv:gr-qc/9307038 [gr-qc]].

\bibitem{Iyer:1994ys}
V.~Iyer and R.~M.~Wald,
``Some properties of Noether charge and a proposal for dynamical black hole entropy,''
Phys. Rev. D \textbf{50}, 846-864 (1994)
[arXiv:gr-qc/9403028 [gr-qc]].

\bibitem{Wald:1999wa}
R.~M.~Wald and A.~Zoupas,
``A General definition of 'conserved quantities' in general relativity and other theories of gravity,''
Phys. Rev. D \textbf{61}, 084027 (2000)
[arXiv:gr-qc/9911095 [gr-qc]].

\bibitem{Hammad:2019oyb}
F.~Hammad, D.~Dijamco, A.~Torres-Rivas and D.~B\'erub\'e,
``Noether charge and black hole entropy in teleparallel gravity,''
Phys. Rev. D \textbf{100}, no.12, 124040 (2019)
[arXiv:1912.08811 [gr-qc]].

\bibitem{Emtsova:2019moq}
E.~D.~Emtsova, A.~N.~Petrov and A.~V.~Toporensky,
``Conserved currents and superpotentials in teleparallel equivalent of GR,''
Class. Quant. Grav. \textbf{37}, no.9, 095006 (2020)
[arXiv:1910.08960 [gr-qc]].


\bibitem{Herrera:2017ztd}
F.~Herrera and Y.~V\'asquez,
``AdS and Lifshitz black hole solutions in conformal gravity sourced with a scalar field,''
Phys. Lett. B \textbf{782}, 305-315 (2018)
[arXiv:1711.07015 [gr-qc]].

\bibitem{Dengiz:2020fpe}
S.~Dengiz, E.~Kilicarslan and M.~R.~Setare,
``Lee-Wald Charge and Asymptotic Behaviors of the Weyl-invariant Topologically Massive Gravity,''
Class. Quant. Grav. \textbf{37}, no.21, 215016 (2020)
[arXiv:2002.00345 [hep-th]].


\bibitem{Anastasiou:2021tlv}
G.~Anastasiou, I.~J.~Araya, C.~Corral and R.~Olea,
``Noether-Wald charges in six-dimensional Critical Gravity,''
JHEP \textbf{07}, 156 (2021)
[arXiv:2105.02924 [hep-th]].

\bibitem{Gecse:2008hj}
Z.~Gecse and S.~Khlebnikov,
``Classical thermodynamics of gravitational collapse,''
Phys. Rev. D \textbf{77} (2008), 104003
[arXiv:0801.3662 [hep-th]].


\bibitem{Constantineau:2011fw}
B.~Constantineau and A.~Edery,
``Numerical thermodynamic studies of classical gravitational collapse in 3+1 and 4+1 dimensions,''
Phys. Rev. D \textbf{84} (2011), 084032
[arXiv:1103.5272 [gr-qc]].


\bibitem{Martinez:2010ti}
C.~Martinez and A.~Montecinos,
``Phase transitions in charged topological black holes dressed with a scalar hair,''
Phys. Rev. D \textbf{82} (2010), 127501
[arXiv:1009.5681 [hep-th]].



\bibitem{Gonzalez:2013aca}
P.~A.~Gonz\'alez, E.~Papantonopoulos, J.~Saavedra and Y.~V\'asquez,
``Four-Dimensional Asymptotically AdS Black Holes with Scalar Hair,''
JHEP \textbf{12} (2013), 021
[arXiv:1309.2161 [gr-qc]].




\bibitem{Gonzalez:2014tga}
P.~A.~Gonz\'alez, E.~Papantonopoulos, J.~Saavedra and Y.~V\'asquez,
``Extremal Hairy Black Holes,''
JHEP \textbf{11} (2014), 011
[arXiv:1408.7009 [gr-qc]].





\bibitem{Lagos:2020oek}
M.~Lagos, P.~G.~Ferreira and O.~J.~Tattersall,
``Anomalous decay rate of quasinormal modes,''
Phys. Rev. D \textbf{101} (2020) no.8, 084018
[arXiv:2002.01897 [gr-qc]].


\bibitem{Aragon:2020tvq}
A.~Arag\'on, P.~A.~Gonz\'alez, E.~Papantonopoulos and Y.~V\'asquez,
``Anomalous decay rate of quasinormal modes in Schwarzschild-dS and Schwarzschild-AdS black holes,''
JHEP \textbf{08} (2020), 120
[arXiv:2004.09386 [gr-qc]].



\bibitem{Perlick:2021aok}
V.~Perlick and O.~Y.~Tsupko,
``Calculating black hole shadows: Review of analytical studies,''
Phys. Rept. \textbf{947}, 1-39 (2022)
[arXiv:2105.07101 [gr-qc]].

\bibitem{Wang:2024lte}
X.~J.~Wang, Y.~Meng, X.~M.~Kuang and K.~Liao,
``Distinguishing black holes with and without spontaneous scalarization in Einstein-scalar-Gauss\textendash{}Bonnet theories via optical features,''
Eur. Phys. J. C \textbf{84}, no.12, 1243 (2024)
[arXiv:2409.20200 [gr-qc]].


\bibitem{EventHorizonTelescope:2019dse}
K.~Akiyama \textit{et al.} [Event Horizon Telescope],
``First M87 Event Horizon Telescope Results. I. The Shadow of the Supermassive Black Hole,''
Astrophys. J. Lett. \textbf{875}, L1 (2019)
[arXiv:1906.11238 [astro-ph.GA]].






\bibitem{EventHorizonTelescope:2021dqv}
P.~Kocherlakota \textit{et al.} [Event Horizon Telescope],
``Constraints on black-hole charges with the 2017 EHT observations of M87*,''
Phys. Rev. D \textbf{103}, no.10, 104047 (2021)
[arXiv:2105.09343 [gr-qc]].


\bibitem{EventHorizonTelescope:2022wkp}
K.~Akiyama \textit{et al.} [Event Horizon Telescope],
``First Sagittarius A* Event Horizon Telescope Results. I. The Shadow of the Supermassive Black Hole in the Center of the Milky Way,''
Astrophys. J. Lett. \textbf{930}, no.2, L12 (2022)
[arXiv:2311.08680 [astro-ph.HE]].


\bibitem{EventHorizonTelescope:2022xqj}
K.~Akiyama \textit{et al.} [Event Horizon Telescope],
``First Sagittarius A* Event Horizon Telescope Results. VI. Testing the Black Hole Metric,''
Astrophys. J. Lett. \textbf{930}, no.2, L17 (2022)
[arXiv:2311.09484 [astro-ph.HE]].





  

\end{thebibliography}
\end{document}